\documentclass[%
 reprint,
superscriptaddress,
nofootinbib,
 amsmath,amssymb,
 aps,
longbibliography,
floats,
floatfix,
]{revtex4-2}

\usepackage[utf8]{inputenc} % allow utf-8 input
\usepackage[T1]{fontenc}    % use 8-bit T1 fonts
\usepackage[colorlinks=true,urlcolor=blue,linkcolor=blue,citecolor=blue]{hyperref}       % hyperlinks
\usepackage{url}            % simple URL typesetting
\usepackage{booktabs}       % professional-quality tables
\usepackage{amsfonts}       % blackboard math symbols
\usepackage{nicefrac}       % compact symbols for 1/2, etc.
\usepackage{microtype}      % microtypography
\usepackage{graphicx}
\usepackage{wrapfig}
\usepackage[toc,page]{appendix}
\usepackage{subcaption}
\usepackage{todonotes}
\usepackage{amsmath}

\newcommand{\ndive}{\texttt{NDIVE}}
\newcommand{\fndive}{\texttt{FTAG+NDIVE}}
\newcommand{\pt}{$p_{T}$}

\newcommand{\jax}{\texttt{JAX}}

\begin{document}

\preprint{APS/123-QED}

\title{Differentiable Vertex Fitting for Jet Flavour Tagging}

\author{Rachel E. C. Smith}
\email[Corresponding authors: \href{mailto:recsmith@slac.stanford.edu,miochoa@lip.pt,makagan@slac.stanford.edu}{recsmith@slac.stanford.edu, \\ miochoa@lip.pt, makagan@slac.stanford.edu}]{}
\affiliation{SLAC National Accelerator Laboratory}
\author{In{\^e}s Ochoa}
\email[Corresponding authors: \href{mailto:recsmith@slac.stanford.edu,miochoa@lip.pt,makagan@slac.stanford.edu}{recsmith@slac.stanford.edu, \\ miochoa@lip.pt, makagan@slac.stanford.edu}]{}
\affiliation{Laboratory of Instrumentation and Experimental Particle Physics, Lisbon}
\author{R{\'u}ben In{\'a}cio}
\affiliation{Laboratory of Instrumentation and Experimental Particle Physics, Lisbon}
\author{Jonathan Shoemaker}
\affiliation{SLAC National Accelerator Laboratory}
\author{Michael Kagan}
\email[Corresponding authors: \href{mailto:recsmith@slac.stanford.edu,miochoa@lip.pt,makagan@slac.stanford.edu}{recsmith@slac.stanford.edu, \\ miochoa@lip.pt, makagan@slac.stanford.edu}]{}
\affiliation{SLAC National Accelerator Laboratory}

\begin{abstract}
   \noindent We propose a differentiable vertex fitting algorithm that can be used for secondary vertex fitting, and that can be seamlessly integrated into neural networks for jet flavour tagging. Vertex fitting is formulated as an optimization problem where gradients of the optimized solution vertex are defined through implicit differentiation and can be passed to upstream or downstream neural network components for network training.  
   More broadly, this is an application of differentiable programming to integrate physics knowledge into neural network models in high energy physics. We demonstrate how differentiable secondary vertex fitting can be integrated into larger transformer-based models for flavour tagging and improve heavy flavour jet classification.
\end{abstract}

\maketitle

\section{Introduction}

Flavour tagging, the identification of jets containing hadrons with heavy flavour bottom and charm quarks (referred to as $b$-tagging and $c$-tagging, respectively), is an essential task for studying a wide array of physical processes at the Large Hadron Collider (LHC)~\cite{Evans_2008} and other particle colliders. Flavour tagging relies on the unique properties of heavy quark hadrons, including: (a) the presence of a secondary vertex (SV) displaced from the primary collision owing to the long lifetime of heavy flavour hadrons before they decay, (b) hard fragmentation, (c) leptons from the decay of the heavy flavour hadron. As such, classic flavour tagging algorithms have relied on reconstructing secondary vertices (vertex-based flavour tagging), identifying particle trajectories (also called tracks) with large displacement, or impact parameter, with respect to the primary collision point (track-based flavour tagging), or finding leptons inside of jets (lepton-based flavour tagging)~\cite{atlasbtag_2016,CMSbtag_2013,Sirunyan_2018,atlasbtag_2019,atlasbtag_2023}.

Recent incarnations of flavour tagging algorithms use modern neural networks~\cite{ATL-PHYS-PUB-2017-003,PhysRevD.94.112002,CMS-DP-2018-058,Bols_2020,ATL-PHYS-PUB-2020-014,ATL-PHYS-PUB-2022-027,ATL-PLOT-FTAG-2023-01,ATL-PHYS-PUB-2023-021,Bols_2020,CMS-DP-2020-002,PhysRevD.102.012010,Shlomi_2021,ATL-PHYS-PUB-2022-027,GOTO2023167836}, such as graph neural networks~\cite{DBLP:journals/corr/abs-2104-13478} or transformers~\cite{NIPS2017_3f5ee243}, to process features from the set of particles within a jet and subsequently perform classification. In some cases, information from secondary vertices, fit by classic algorithms, are used as input to neural flavour tagging models because they are a key signature of heavy flavour jets. However, neural flavour tagging models do not explicitly fit secondary vertices. In this paper we present \ndive\  (``en-dive''), the first differentiable vertex fitting  algorithm that can be integrated naturally into neural networks for tasks such as flavour tagging. We also present a novel flavour tagging model that integrates \ndive\ within a larger neural network to improve flavour tagging performance, which we denote \fndive. 

In \ndive, we consider a least squares optimization formulation of vertex fitting: an inclusive vertex, i.e. a single vertex (as opposed to multiple vertices), is fit  via minimization of a weighted $\chi^2$~\cite{BILLOIR1985115,Fruhwirth:2007hz} over a set of particle tracks. Once a vertex solution is found, the gradient of the solution vertex with respect to input weights or particle features is defined using implicit differentiation, thus enabling \ndive\ to pass gradients to upstream or downstream functions for gradient-based optimization, such as in neural network training. 
\ndive\ provides a differentiable programming based scheme to integrate physics knowledge, such as the geometry of particle trajectories, into neural networks.

Related work is discussed in Sec.~\ref{sec:related}. An overview of the methods used in the \ndive\ algorithm is found in Sec.~\ref{sec:diva}.  The implementations of neural network models using \ndive\ are found in Sec.~\ref{sec:models}. Sec.~\ref{sec:exp} presents experiments. 
\begin{figure}[t]
\centering
    \begin{subfigure}[b]{0.49\textwidth}
    \centering
    \includegraphics[width=\textwidth]{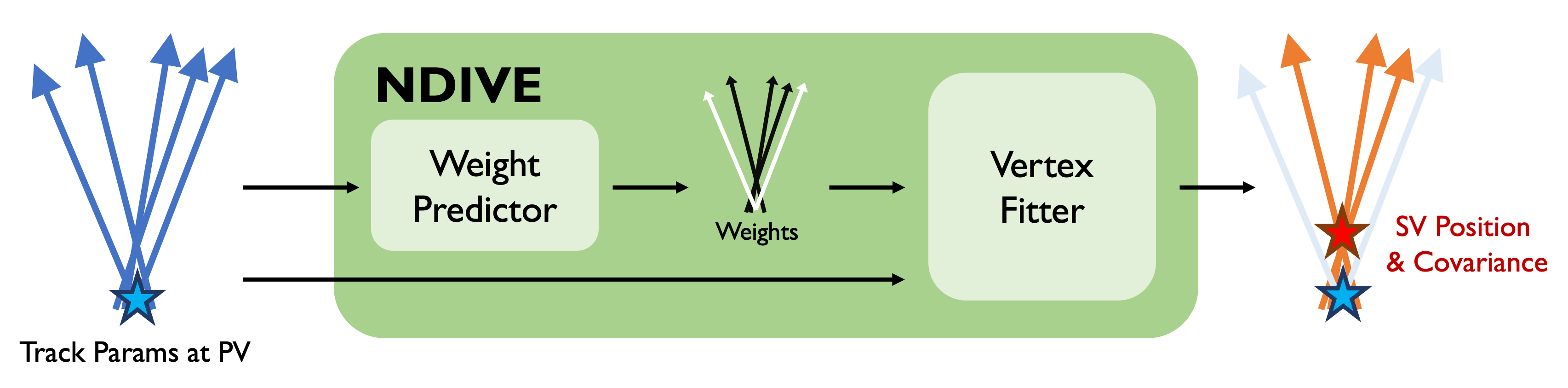}
    \end{subfigure}
\caption{Illustration of the \ndive\ module. PV stands for primary vertex. SV stands for secondary vertex, or, more generally, the vertex predicted by the weighted fit.}
\label{fig:ndive}
\end{figure}

\vspace{0.1cm}
\noindent \textit{Contributions:}
\begin{itemize}
\item We connect vertex fitting to the concept of \textit{optimization as a layer}. We show how implicit differentiation can be used to enable differentiation through the solutions of the optimized vertex fits. Combined together, these concepts allow us to introduce the differentiable vertex fitting algorithm \ndive\footnote{Repo: \href{https://github.com/rachsmith1/NDIVE}{https://github.com/rachsmith1/NDIVE}}.

\item We develop a strategy for integrating \ndive\ into machine learning (ML) models such that the fitted vertex can be integrated into further neural processing layers. As such, we propose a new model architecture for flavour tagging, \fndive, that directly utilizes \ndive\ and improves heavy flavour jet classification performance.

\item We provide performance studies of both the vertex fitting capabilities of \ndive\ alone as well as its integration into larger models. 

\end{itemize}

\section{Related Work}\label{sec:related}

A wide array of \textit{classic} (i.e. not ML based) algorithms have been developed for secondary vertex fitting and jet flavour tagging, including algorithms based on identifying tracks with large impact parameters, identifying leptons within jets, and inclusive and multi-vertex reconstruction algorithms (for an overview, see e.g. references~\cite{Fruhwirth2021} and~\cite{Giacinto2010}).  Early usage of ML for flavour tagging focused on combining high-level features such as jet features, event features, and features derived from the output of classic algorithms (see e.g.~\cite{BORTOLOTTO1991459} for an early example, and more recent usage at the LHC~\cite{atlasbtag_2016,CMSbtag_2013,Sirunyan_2018,atlasbtag_2019,atlasbtag_2023}). More recently, rather than first deriving high-level features, deep neural networks for flavour tagging have been employed to process variable length sets of track features within jets, e.g. using recurrent neural networks~\cite{ATL-PHYS-PUB-2017-003,PhysRevD.94.112002,CMS-DP-2018-058,Bols_2020}, deep sets~\cite{ATL-PHYS-PUB-2020-014}, graph neural networks~\cite{ATL-PHYS-PUB-2022-027,PhysRevD.101.056019}, and transformers~\cite{ATL-PHYS-PUB-2023-021,DBLP:conf/icml/QuLQ22}. In some cases, sets of features from secondary vertices fit with classic algorithms are also processed within the model~\cite{Bols_2020,CMS-DP-2020-002,PhysRevD.102.012010}. Some models have begun to explore vertex finding~\cite{Shlomi_2021,ATL-PHYS-PUB-2022-027,GOTO2023167836}, i.e. the identification of tracks that are predicted to belong to a secondary vertex, but do so without performing explicit vertex fitting or explicitly using the geometric information contained in track trajectories. Our work can be seen as an extension of track-based models that now explicitly integrates vertex fitting and track geometry, using differentiable versions of classic algorithms, into the model architecture and predictions.

Our models rely on the use of differentiable optimization as a layer~\cite{10.5555/3305381.3305396,NEURIPS2019_9ce3c52f,cvxpy}. These tools are part of the larger body of work on deep implicit layers (for an overview, see~\cite{impl_tutorial}) that uses implicit differentiation~\cite{Krantz2013,NEURIPS2022_228b9279} to compute gradients of solutions of implicit functions, optimizations, or differential equations. 

Our work can be seen as part of the development in HEP of differentiable programming (DP), i.e.  use of automatic differentiation~\cite{autodiff,JMLR:v18:17-468} (AD) and other gradient estimation techniques to develop differentiable software that can be optimized for a given task and can be integrated with neural networks to extend deep learning to novel computations. AD is the backbone of ML / DP frameworks like \texttt{TensorFlow}~\cite{tensorflow2015-whitepaper}, \jax~\cite{jax2018github}, and \texttt{PyTorch}~\cite{NEURIPS2019_9015pytorch}. Within HEP, AD has been used in histogram fitting in \texttt{pyhf}~\cite{pyhf, pyhf_joss}, in analysis optimization within \texttt{Neos}~\cite{lukas_heinrich_2020_3697981} that also makes use of implicit differentiation and differentiable optimization, in modeling parton distribution functions used by matrix element generators~\cite{pdflow2021,ball2021opensource}, in developing differentiable branching processes and particle showers~\cite{kagan2023branches}, in progress towards differentiable parton showers~\cite{nachman2022morphing}, and in developing differentiable matrix element codes~\cite{Heinrich_2023}.

\section{Methods}\label{sec:diva}

\subsection{Differentiable Optimization}\label{subsec:diffopt}

In differentiable optimization, we both optimize an objective function and derive derivative relations at the optimized point. We denote the objective function $\mathcal{S}(\mathbf{x}, \alpha)$ where $\mathbf{x}$ is the value we want to optimize and $\alpha$ is a set of parameters.  We can thus define the solution of the optimization problem:
\begin{equation}\label{eq:opt}
    \hat{\mathbf{x}}(\alpha) = \arg \min_{\mathbf{x}} \mathcal{S}( \mathbf{x}, \alpha)
\end{equation}

Importantly, we have written out the dependence of the solution on the input, $\hat{\mathbf{x}}(\alpha)$, indicating that a change in $\alpha$ could lead to a different fit solution. Note that the $\alpha$ quantities can depend on upstream parameters. For instance, \ndive\ will use per-track weights predicted by a neural network to determine how much a track contributes to the vertex fit, and thus the derivative of a fit vertex with respect to these weights is needed to train the upstream neural network. More broadly, these derivatives will be needed for training the models in Sec.~\ref{subsec:neural_vtx} and~\ref{subsec:flav_vtx}.

Assuming the objective is continuously differentiable with non-singular Jacobian, the implicit function theorem tells us that we may take derivatives of the fit values with respect to the parameters.
To derive the derivatives of the fit value $\hat{\mathbf{x}}$ with respect to the parameters $\alpha$, we note that $\mathcal{S}(\mathbf{x}, \alpha)$ is at a minimum with respect to $\mathbf{x}$ after optimization, i.e. $\partial_{\mathbf{x}} \mathcal{S}(\mathbf{x}, \alpha) = 0$ when evaluated at $\hat{\mathbf{x}}(\alpha)$. To simplify notation, we will denote  $\hat{\mathcal{G}} \equiv \partial_{\mathbf{x}} \mathcal{S}(\hat{\mathbf{x}}, \alpha)$. Taking then the derivative with respect to $\alpha$ and accounting for the implicit dependence of $\hat{\mathbf{x}}$ on $\alpha$:
\begin{equation}
    \nonumber 0 = \frac{d}{d\alpha} \hat{\mathcal{G}} = \frac{\partial \hat{\mathcal{G}}}{\partial \alpha} + \frac{\partial \hat{\mathcal{G}}}{\partial \mathbf{x}} \frac{\partial \mathbf{x}}{\partial \alpha}
\end{equation}
Rearranging we find:
\begin{equation}\label{eq:impl_deriv}
    \frac{\partial \mathbf{x}}{\partial \alpha} =  -\Big(\frac{\partial \hat{\mathcal{G}}}{\partial \mathbf{x}}\Big)^{-1} \frac{\partial \hat{\mathcal{G}}}{\partial \alpha} 
\end{equation}
Thus Eqn.~\ref{eq:impl_deriv} defines the derivatives of the optimization solution with respect to the objective parameters that may be needed for upstream or downstream components (e.g. for training).

This optimization can be implemented as a neural network layer and fully integrated into AD frameworks, where inputs are the values $\alpha$ and the output is the fit value $\hat{\mathbf{x}}$. In the forward pass, the layer can use a numerical optimization procedure to minimize the objective and find the best fit values. In the backward pass used to compute gradients for backwards mode AD, one can implement a custom derivative definition following Eqn.~\ref{eq:impl_deriv}. One could, alternatively, use standard backpropagation through the numerical solver, which likely optimizes the objective using a iterative solver. However, backgropagating through iterative solvers is both memory inefficient and often prone to numerical instabilities.

\subsection{Inclusive Vertex Fit Formulation}\label{subsec:Billoir}

In vertex fitting, our goal is to find a common point of production for a set of tracks. Heuristically, this can be achieved by finding a space point, the vertex, and a set of momenta for each track at that position. The difference is minimized between the measured track parameters (e.g. perigee parameters with respect to the primary vertex) and the track parameters calculated by extrapolating from the predicted vertex using the predicted set of track momenta to their perigee at the primary vertex. One can thus setup vertex fitting as an optimization problem, where an objective measures the difference between measured and predicted track parameters, and minimizes this difference by finding the optimal vertex point and set of track momenta at that point. Many common vertex fitting algorithms can be formulated as optimizing an objective, and in this work we will focus on the Billoir Algorithm~\cite{BILLOIR1985115} for inclusive vertex fitting that minimizes a least squares objective. 

The values we will optimize are the vertex position, $\mathbf{v}$, and the track momentum at the vertex, $\{\mathbf{p}_i\}$; thus $\mathbf{x} = (\mathbf{v}, \{\mathbf{p}_i\})$. The objective also takes as input data the track perigee parameters, measured with respect to the primary vertex, $\mathbf{q}_i$, and their covariance matrices, $\mathbf{V}_i$, from a set of $N$ tracks (where $N$ can vary for each jet). We will use a weighted formulation of vertex fitting, and thus the objective will also take as input a set of per-track weights, $w_i$, which determine how much each track contributes to the vertex fit. In relation to Sec.~\ref{subsec:diffopt}, we here have $\alpha = \{w_i, \mathbf{q}_i, \mathbf{V}_i\}$.

In this strategy, a least square (LS) objective $\mathcal{S}(\mathbf{v}, \{\mathbf{p}_i\} | \{w_i, \mathbf{q}_i, \mathbf{V}_i\} )$ is minimized. Denoting the objective as $\mathcal{S}$ for ease of notation, the LS objective is:
\begin{equation}\label{eq:ls_obj}
        \mathcal{S} = \sum_{i=1}^{N} w_i (\mathbf{q}_i - \mathbf{h}_i(\mathbf{v}, \mathbf{p}_i))^{T} \mathbf{V}_i^{-1} (\mathbf{q}_i - \mathbf{h}_i(\mathbf{v}, \mathbf{p}_i)),
\end{equation}
where $\mathbf{h}_i(\mathbf{v}, \mathbf{p}_i)=\mathbf{q}_{\mathrm{model}, i}$ is a given track model (which can be nonlinear, e.g. a helical model of a curved track) that relates the predicted track parameters to the fit vertex and momenta parameters. In this work, the track models $\mathbf{h}_i (\cdot)$ do not include material interactions and assumes a uniform magnetic field. 
As noted above, equation~\ref{eq:ls_obj} includes a set of per-track weights $\{w_i\}$ that are used as an effective track selection (i.e. they may be the output of a vertex finding step). Later in this paper, we will estimate these probabilities with a neural network.  

When a non-linear track model is used, this fit takes the form of a non-linear regression. To optimize this objective, an iterative linearized optimization is performed. Given an initial guess for the values of $(\mathbf{v}^{(0)}, \{\mathbf{p}^{(0)}_{i}\})$, a linear approximation in both $\mathbf{v}$ and $\{\mathbf{p}_i\}$ of the track model $\mathbf{h}_i(\cdot)$ is obtained using the perigee representation of the tracks. The resulting linearized LS objective optimization can be solved exactly, resulting in new values $\hat{\mathbf{v}}^{(1)}, \{\hat{\mathbf{p}}^{(1)}_i\}$. These fit values can then be used as the initial values in a subsequent iteration of the optimization where the track model is then linearized around the fit values from the previous iteration. In practice we use a fixed 10 iterations but found the fit usually converged in 2 to 4 iterations for the models described in Sec.~\ref{sec:models}. A detailed derivation of a single iteration of the fit can be found in App.~\ref{app:vertexfit}.

\subsection{Differentiable Track Extrapolation}\label{subsec:Extrapolation}

Measured track parameters are typically expressed at a given point along the trajectory with respect to a reference point (a common choice is the perigee parameters with respect to the primary vertex). Track extrapolation refers to the propagation of track parameters, using the track model, to a specified point along a trajectory. Thus, a track extrapolator is a transformation ($T$) of initial track perigee parameters ($q_0$) from a given reference point ($P_0$) to a new reference point ($P^\prime$), e.g. $q^\prime = T(q_0, P_0, P^\prime)$. By incorporating the transformation into a differentiable programming framework, e.g. \texttt{JAX}, a differentiable track extrapolator module can readily be used in a neural network.

Within our models, the Track Extrapolator module can incorporate our knowledge of the expected track geometry to extrapolate each track into the point of closest approach to the newly estimated vertex. Measured tracks are initially represented by their perigee parameterization with respect to the primary vertex of the collision. The Track Extrapolator module allows us to construct an alternative representation for each track, where its parameters are calculated using the vertex estimated by \ndive\ as reference point, $\mathbf{q^\prime}_i$. Further details in how the track trajectories are parameterised and their alternative representations calculated are provided in App.~\ref{app:extrap}. 

\section{Models}\label{sec:models}

\subsection{\ndive: Bringing It All Together}\label{subsec:neural_vtx}

We formulate \ndive\ as a transformer-guided vertex fitting model that takes as input track- and jet-level kinematic variables. Jet-level features are concatenated onto all tracks. These variables are fed into a transformer which outputs a set of per-track weights each between 0 and 1. These weights, along with the measured track perigee parameters and their covariance matrix, are then fed to a differentiable vertex fitting layer. In the forward pass, this layer performs a weighted formulation of the iterative Billoir vertex fitting algorithm, as described in Sec.~\ref{subsec:Billoir}, outputting a three-dimensional prediction for the vertex position, the per-track momenta at the predicted vertex, and the associated covariance matrix. The loss function is the mean absolute error between the predicted vertex position and the true secondary vertex position. In the backward pass, gradients are computed via implicit differentiation as described in Sec.~\ref{subsec:diffopt} using the LS objective in Eqn.~\ref{eq:ls_obj}. The \ndive\ model is illustrated in Fig.~\ref{fig:ndive}. 

\subsection{\fndive: Vertex Integrated Flavour Tagging}\label{subsec:flav_vtx}

To perform flavour tagging, \ndive\ must be integrated with other neural layers to perform classification. There are many possible strategies for this integration, and we present a single strategy here. In \fndive, we choose a baseline architecture that is similar to models already in use for flavour tagging and integrate \ndive\ as an additional module in the network to generate new features (e.g. the vertex information) and to guide the exploration of the vertex geometry (e.g. through estimating track parameters after track extrapolation to the fit vertex).   We leave exploration of other strategies as further work.

\vspace{0.2cm}
\noindent\textbf{Baseline Architecture:} We base the architecture for our flavour tagging model off the recently proposed \texttt{GN2} architecture developed by ATLAS~\cite{ATL-PHYS-PUB-2022-027,ATL-PLOT-FTAG-2023-01}, albeit with some modifications, mostly concerning the number of model layers, due to the relatively smaller size of the training dataset.  
\begin{figure}[ht!]
\centering
    \begin{subfigure}[b]{0.4\textwidth}
    \centering
    \includegraphics[width=\textwidth]{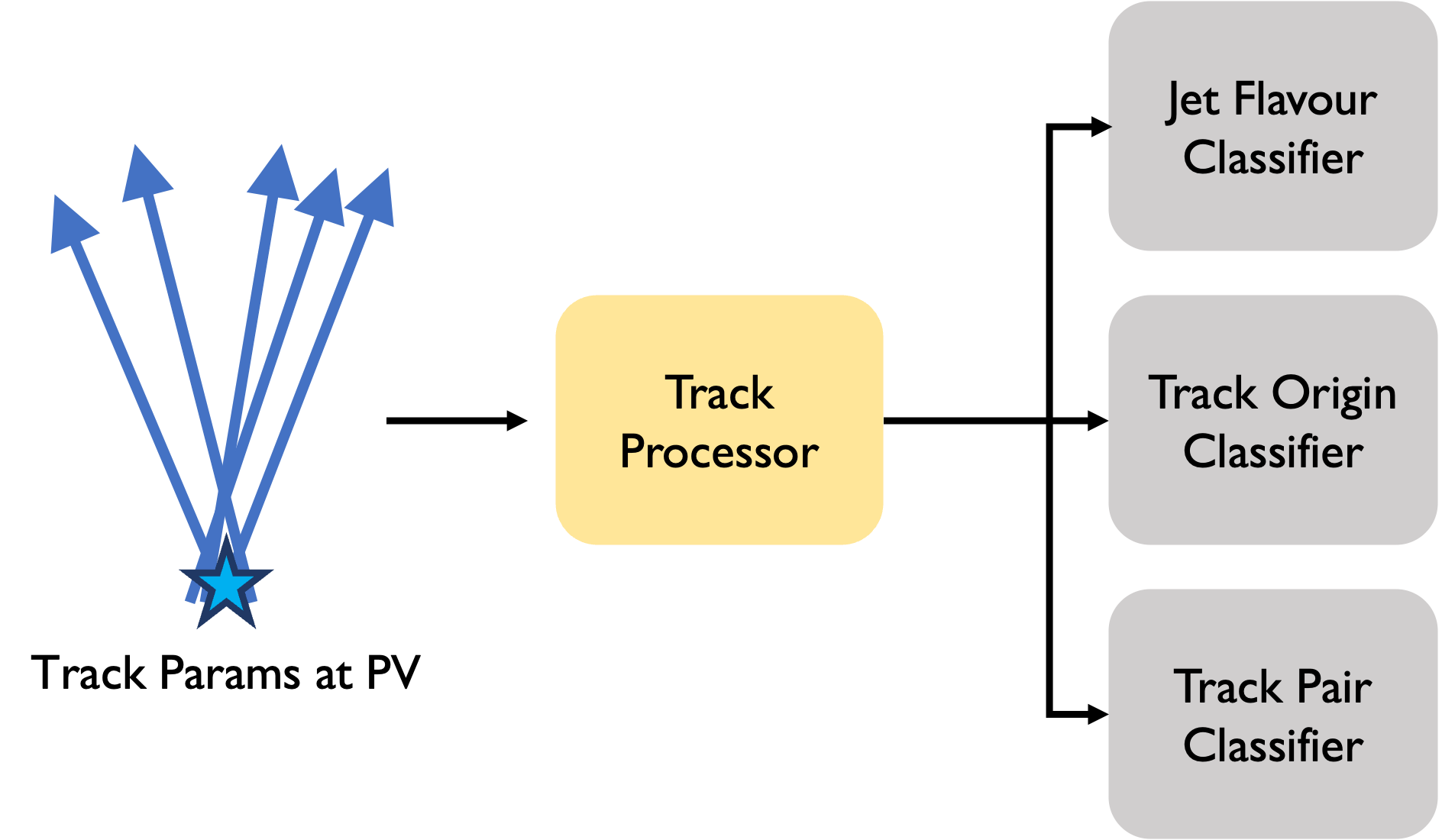}
    \end{subfigure}
\caption{Illustration of the baseline architecture of the flavour tagging model.}
\label{fig:baseline}
\end{figure}

Track- and jet-level kinematic variables are inputs to the model. One additional feature used in our model is the primary vertex position, which is appended to the features for each track and is used to help the model identify differences in track features when represented at different reference points. These variables are fed to a ``track processor'' transformer network. The output features of the track processor constitute a representation of each track in the jet, conditional on the other tracks. These outputs are combined via global attention pooling~\cite{DBLP:journals/corr/LiTBZ15} in a weighted sum, where the weights are learned during training, to construct a global jet representation. Both the track- and jet-level representations are used in the subsequent training objectives.

As in \texttt{GN2}, in addition to the jet flavour classification training objective, the model has two other auxiliary training objectives, one of which is a prediction of the track origin (from a $b$- or $c$-hadron, from the primary vertex, or from some other origin) and the other is a prediction of track-pair compatibility. This second auxiliary training objective effectively performs vertex finding, however, as noted earlier, it does not explicitly integrate track geometry and vertex fitting. As such, vertex finding in this form can not learn from the quality of a downstream vertex fit. The baseline architecture is illustrated in Fig.~\ref{fig:baseline}. Each auxiliary task has an associated cross entropy loss.

This baseline architecture is used for comparison between models that do not use differentiable vertex fitting and those that do, namely by integrating \ndive\ into the track processing step prior to the training objectives. This \ndive-integrated architecture is described below.

\begin{figure}[ht!]
\centering
    \begin{subfigure}[b]{0.49\textwidth}
    \centering
    \includegraphics[width=\textwidth]{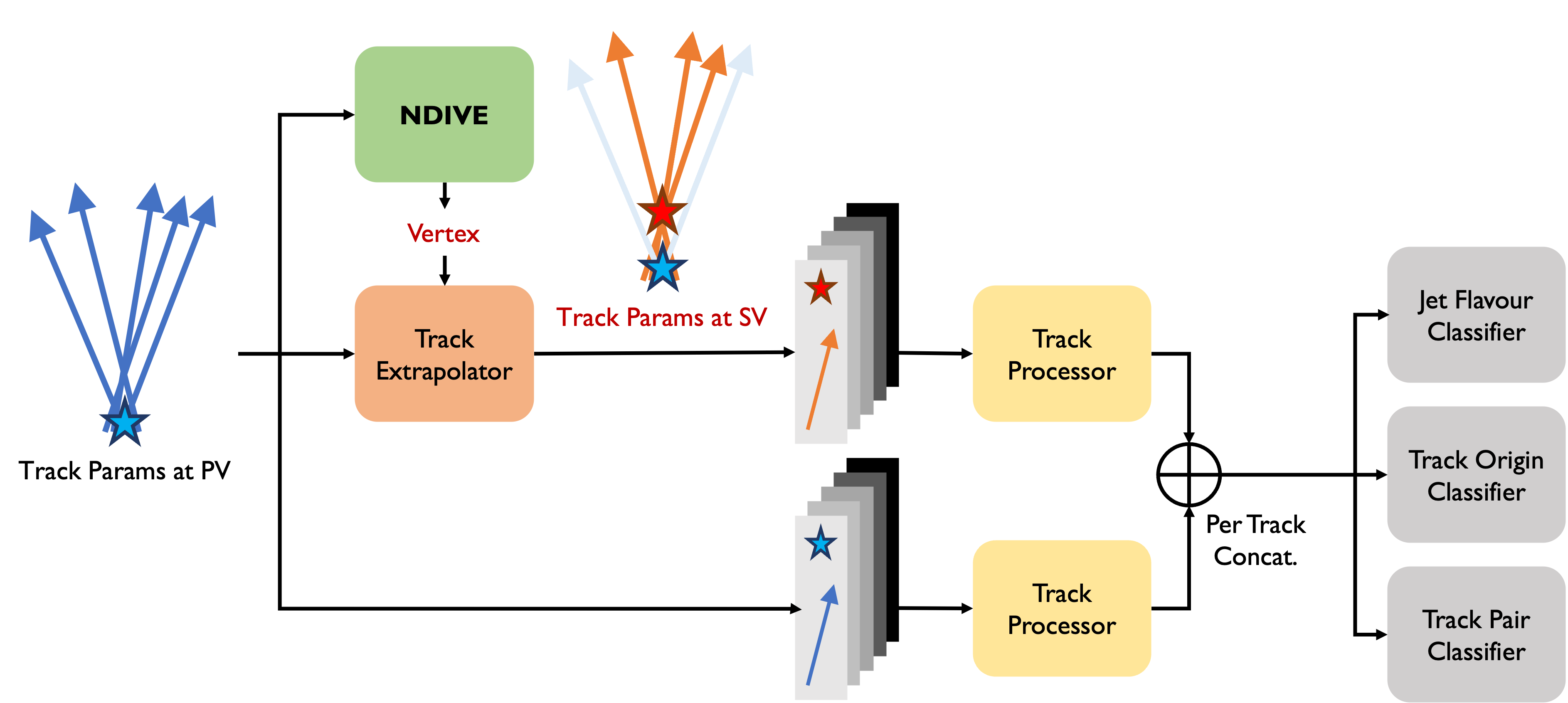}
    \end{subfigure}
\caption{Illustration of the architecture of the \fndive\ vertex integrated flavour tagging model.}
\label{fig:flavtag}
\end{figure}

\vspace{0.2cm}
\noindent \textbf{\ndive\ Integration:}
\ndive\ is integrated into the larger flavour tagging model as an additional step of track processing. The vertex fit predicted by \ndive\ is used as an input to a differentiable track extrapolator, as described in Sec.~\ref{subsec:Extrapolation}. The predicted vertex position and the track perigee parameters calculated with respect to the predicted vertex are then given as inputs to the track processor module. The \textit{same} track processor module also processes the original track perigee parameters calculated with respect to the primary vertex. As noted earlier, the primary vertex position is appended as a feature for tracks parameters calculated with respect to the primary vertex. The outputs of processing both sets of track parameters are concatenated to give the final track representation. As in the baseline model, these track level outputs are combined with a weighted sum to construct a global jet representation. The track and jet representations are then used in the subsequent training objectives. The full \fndive\ vertex integrated model is illustrated in Fig.~\ref{fig:flavtag}. All model components, including the \ndive\ block, the track processor, and the task classifiers, are jointly optimized in \fndive. In addition to baseline losses, the mean absolute error \ndive\ loss is also used in the training.

\section{Experiments}\label{sec:exp}

We start by using \ndive\ for vertex fitting in $b$-, $c$- and light jets and we evaluate its track selection and vertex reconstruction performance as a function of the properties of the jets. Then, we evaluate the \fndive\ vertex integrated flavour tagging model that aims to classify jets containing $b$-hadrons, $c$-hadrons, or jets originating from light quarks, and compare it to the performance achieved by a state-of-the-art tagger with the baseline architecture, as described in Sec.~\ref{subsec:flav_vtx}. 

Both the \ndive\ Weight Predictor and the Track Processor in the baseline architecture / \fndive\ are transformer encoders that consist of two layers. The first layer is a single-headed self-attention mechanism. The second is a fully-connected dense layer. A residual connection~\cite{He_2016_CVPR} is employed around each layer followed by layer normalization~\cite{ba2016layer}. A single transformer layer was used due to the relatively small data set size. The jet flavour classifier, track origin classifier, and track pair classifier each consist of four dense layers. The models were optimized using the Novograd optimizer~\cite{novograd} (which we found provided better training stability than Adam) with a learning rate of 1e-4.  A batch size of 100 jets was used.

\begin{figure*}[ht!]
\centering
    \begin{subfigure}[b]{0.32\textwidth}
    \centering
    \includegraphics[width=\textwidth]{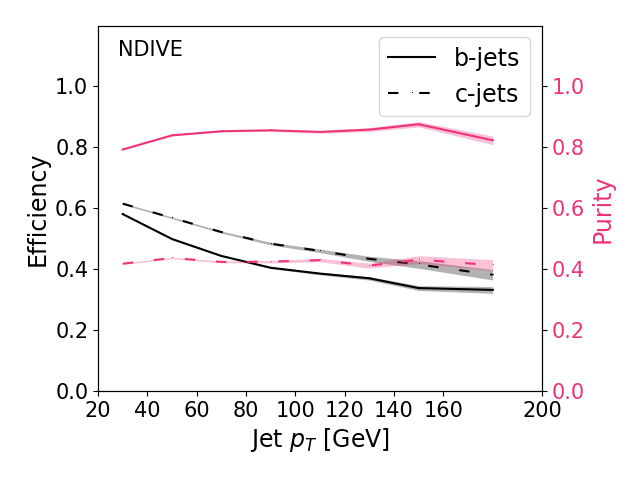}
    \end{subfigure}
    \begin{subfigure}[b]{0.32\textwidth}
    \centering
    \includegraphics[width=\textwidth]{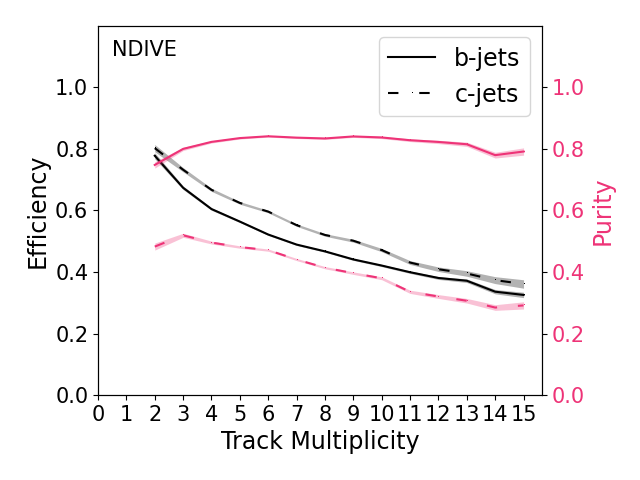}
    \end{subfigure}    
    \begin{subfigure}[b]{0.32\textwidth}
    \centering
    \includegraphics[width=\textwidth]{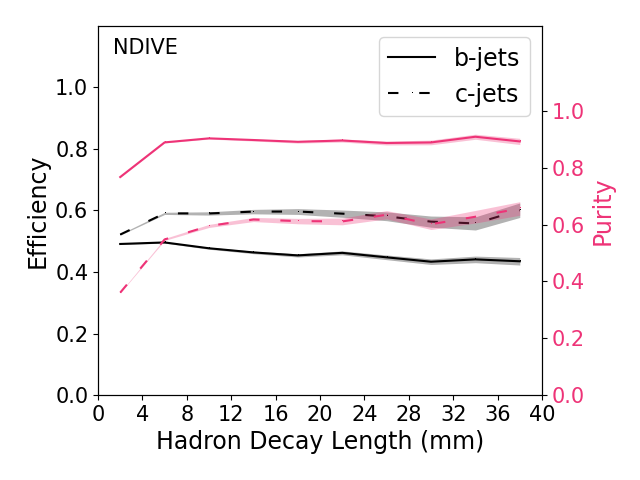}
    \end{subfigure}        
\caption{Track selection efficiencies (left) and purities (right) achieved by \ndive\ with a selection threshold of 0.5, as a function of the jet \pt, track multiplicity, and hadron decay length, for $b$- and $c$-jets. The error bars correspond to the square root of the variance of a binomial distribution.}
\label{fig:ndive_eff_purity}
\end{figure*}

\subsection{Data Samples}
We use the simulated dataset from reference~\cite{zenodo_dataset}, generated with Pythia8~\cite{Sjostrand:2007gs}, and a basic detector simulation is performed with Delphes~\cite{deFavereau:2013fsa} to emulate a detector similar to ATLAS. The data set consists of jets from top-pair production events in proton-proton collisions at $\sqrt{s}=14$~TeV. Pileup collisions are not included, i.e. only a single proton-proton collision is present in each event. Jets are constructed using calorimeter energy deposits with the anti-$k_T$ algorithm~\cite{antikt} with radius 0.4. Charged particle tracks are matched to a jet if they are within a cone of 0.4 from its axis. 500 thousand jets are used for training, 180 thousand for validation, and 180 thousand for testing and making the final performance comparisons. The fractions of $b/c/$light jets are $\sim1/3$ of each sample and have the same distributions in jet \pt\ and $\eta$.

The input features to the vertex fitting and flavour tagging models are the jet \pt, jet $\eta$, and the set of tracks associated to the jet. Tracks are specified by their perigee parameterization with respect to the primary vertex, such that each track has features $d_0, z_0, \phi, \theta, \rho$ and an associated diagonal covariance matrix.  Additionally, signed impact parameter significances are defined by assigning a lifetime sign to each track.

\subsection{Performance: \ndive}

The performance on \ndive\ is examined through its track selection ability and the resolution of the fit vertices.

The success of the vertex fitting task depends on the quality of the track selection, i.e. how efficiently tracks from the heavy flavour decay are selected and how often non-heavy flavour decay tracks are selected. The Weight Predictor component of \ndive~predicts a track's importance to the fitted vertex by assigning weights. As every track in a jet is used in the vertex fit, albeit with weights, we interpret the notion of a ``selected track" as a track which contributes significantly to the fit. However, it is important to note that due to the structure of the LS objective (the LS objective can be multiplied by a constant without affecting the optimization result), only the weight of each track in a jet relative to the weights of other tracks in the jet impact the fit. As such, for performance studies, we normalize the track weights to the maximum weight assigned per jet, such that only the relative importance of tracks is taken into account. Based on this normalized weight, an effective track selection can then be implemented to study the overall performance. In this work, we choose to define selected tracks as those with relative weight greater than $0.5$. A different choice for the relative weight threshold could be made, which would merely indicate a different threshold of importance a track will have on a given vertex fit. Note, this normalization is only made for the interpretation for performance studies, and was not implemented in the actual models.

\begin{figure}[ht!]
\centering
    \begin{subfigure}[b]{0.49\textwidth}
    \centering
    \includegraphics[width=0.75\textwidth]{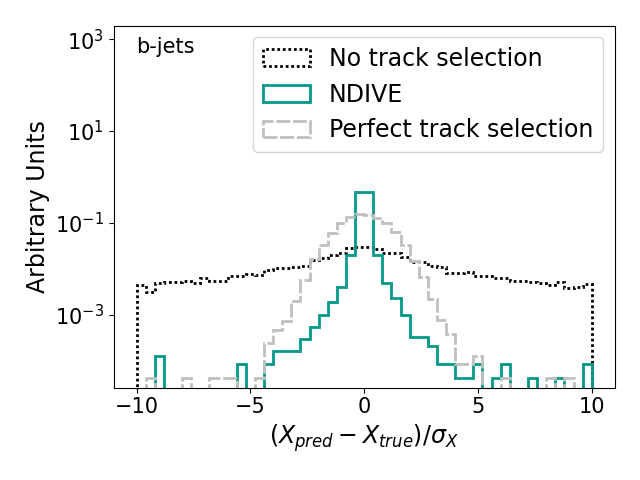}
    \end{subfigure}\\  
\caption{Difference between the fit and true vertex $x$-coordinate divided by the square root of the fit vertex variance, for $b$-jets, comparing \ndive~with vertex fitting with no track selection and with perfect track selection.}
\label{fig:ndive_score_x}
\end{figure}

\begin{figure*}[ht!]
\centering
    \begin{subfigure}[b]{0.32\textwidth}
    \centering
    \includegraphics[width=\textwidth]{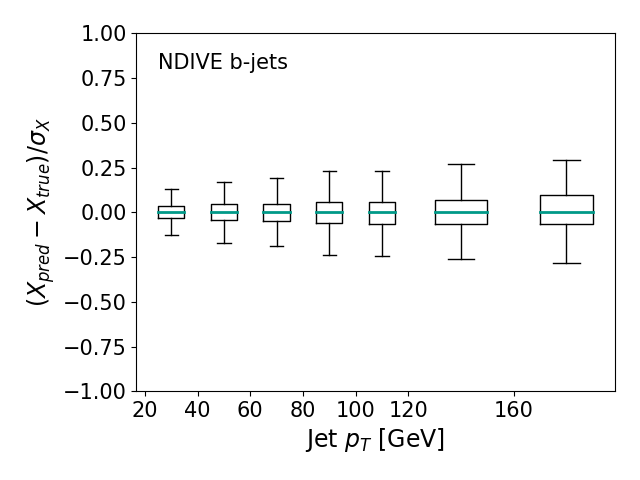}
    \end{subfigure}
    \begin{subfigure}[b]{0.32\textwidth}
    \centering
    \includegraphics[width=\textwidth]{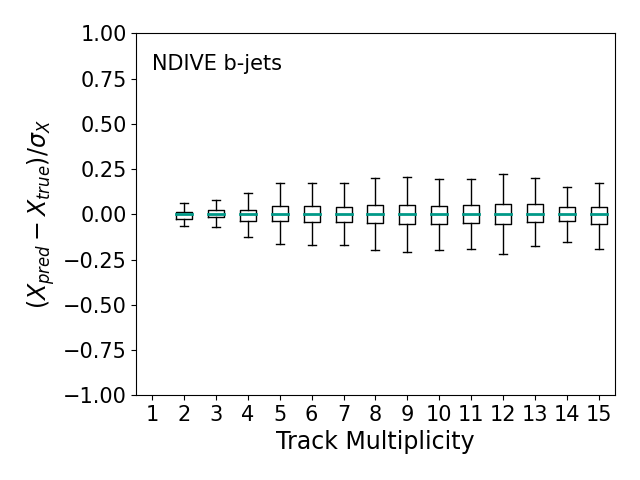}
    \end{subfigure}
    \begin{subfigure}[b]{0.32\textwidth}
    \centering
    \includegraphics[width=\textwidth]{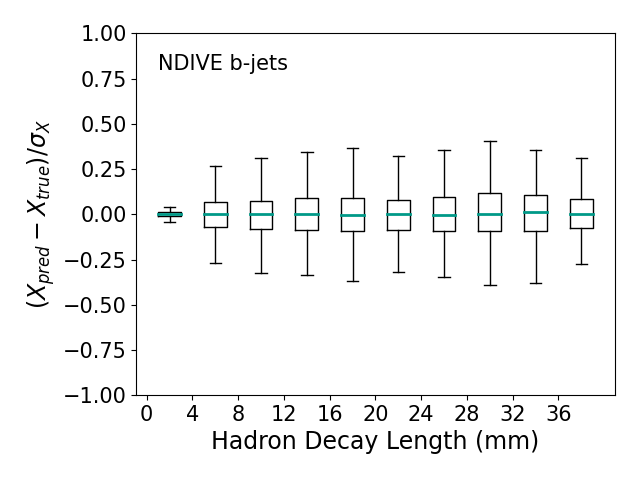}
    \end{subfigure}        
\caption{Median difference between the fit and true vertex $x$-coordinate divided by the square root of the fit variance per-track, for $b$-jets, in bins of jet \pt, track multiplicity, and hadron decay length. The boxes indicate the interquartile range (IQR) of the distributions and the error bars cover data points that fall within $1.5$ times the IQR from the box.}
\label{fig:ndive_med}
\end{figure*}

The efficiency of heavy flavour track selection is defined as the fraction of heavy flavour tracks selected out of all heavy flavour tracks in a jet. The purity is defined as the fraction of heavy flavour tracks selected out of all tracks selected in a jet. The efficiency and purity in $b$-jets and $c$-jets is shown in Figure~\ref{fig:ndive_eff_purity}. Both are found to be relatively stable as a function of the jet \pt, track multiplicity and hadron decay length. Intriguingly, we observe a relatively low selection efficiency for this threshold, approximately $50\%$ for $b$-jets. However, given the expected charged track multiplicity of $b$-decays is five, on average at least two tracks are selected and used to form a vertex. Importantly, we find that the purity of the selection is high and thus even with relatively low selection efficiency a successful fit to the secondary vertex can be performed.  We also note that the track selection efficiency increases when \ndive\ is used within the larger \fndive\ vertex-integrated flavour tagging model, as seen in App.~\ref{app:fndiveperf}.

Figure~\ref{fig:ndive_score_x} shows the difference between the fit vertex and the true vertex $x$-coordinate, divided by the square root of the fit vertex variance, for $b$-jets. The distribution has approximately zero mean, indicating an unbiased fit, and has low variance, indicating good estimates of the vertex position within error. Similar distributions are seen for the $y$- and $z$-coordinates. To benchmark its performance, the \ndive\ fit result is compared to a vertex fit performed with all tracks associated to the jet (\emph{no track selection}) and to a vertex fit where only the tracks originating from the secondary vertex are fit, i.e. where the per-track weights assigned by the neural network are replaced by 0 or 1 according to the true track origin (\emph{perfect track selection}). As expected, using all tracks in the $b$-jet will lead to bad estimates of the vertex coordinates. On the other hand, using the perfect track selection leads to small uncertainties on the predicted coordinates. The corresponding figures for $c$- and light-jets can be found in App.~\ref{app:ndiveperf}. The width of light-jet distribution is significantly smaller than the $b$- and $c$-jet distributions because in light-jets all ``true'' tracks originate from the primary vertex. As such, more tracks are typically  available for fitting the vertex and the vertex fit quality is improved. The two dimensional distributions of true vertex versus fit vertex $x$- and $z$-coordinates can also be found in App.~\ref{app:ndiveperf}.

To quantify the performance of the vertex fitting model, the mean and standard deviation of the distributions of the difference between true vertex and fit vertex $x$-coordinate, divided by the fit error, for $b$-jets is shown in Fig.~\ref{fig:ndive_med} as a function of jet $p_T$, jet track multiplicity, and true hadron decay length. The mean difference as a function of all three variables is seen to be approximately zero, indicating the fits are not biased in a specific phase space. The standard deviations tend to be smallest at low $p_T$ and small hadron decay length where \ndive\ is more likely to (incorrectly) select tracks from the primary vertex to use in the fit and thus drive the uncertainty on the fit toward lower values. The fit is unbiased because the decay length in these settings is likely to be close to zero, and thus consistent with the primary vertex tracks.

\subsection{Performance: \fndive}\label{sec:fndiveperf}

Our primary performance metric for the \fndive\ vertex-integrated flavour tagging algorithm is the receiver operating characteristic (ROC) curve, which shows the $c$- and light-jet rejection (one over the false positive rate) versus the $b$-jet efficiency (the true positive rate for $b$-jets) made by scanning a threshold on a discriminant and examining the set of jets passing each threshold. To benchmark gains in performance, we compare \fndive\ with a flavour tagger built only from the baseline architecture. 

To compute the ROC, a discriminant is needed. We construct our discriminant in the same way as \mbox{ATLAS}~\cite{ATL-PHYS-PUB-2022-027} from the jet-level probabilities $p_b$, $p_c$, $p_l$ outputted from the flavour classifier at the end of the model: 
\begin{equation}
D_{b} = \log \frac{p_b}{(1-f_c) p_l + f_c p_c}
\end{equation}

\begin{figure}[ht!]
\centering
    \begin{subfigure}[b]{0.4\textwidth}
    \centering
    \includegraphics[width=\textwidth]{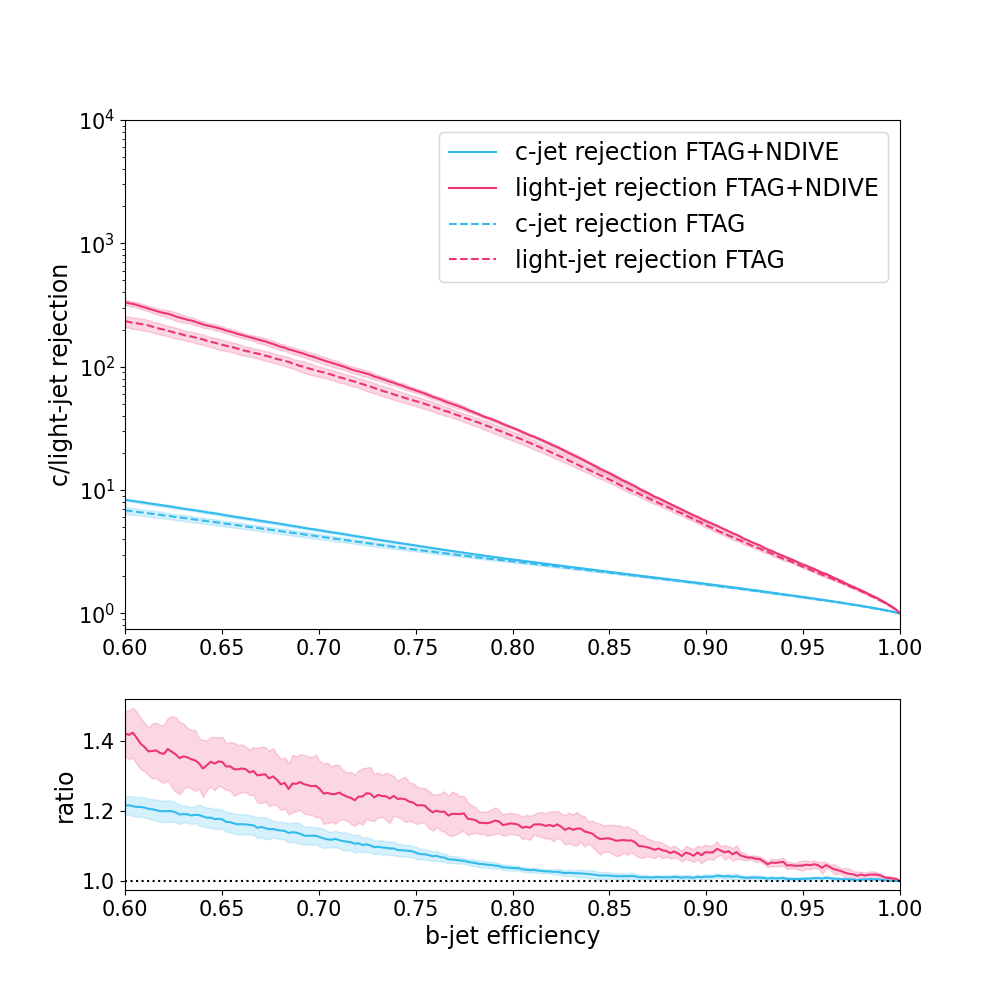}
    \end{subfigure}
\caption{Light-jet and $c$-jet rejection as a function of $b$-tagging efficiency for the flavour tagging models in the baseline architecture and with \ndive\ integration. The error bars are the standard deviation of 5 trainings.}
\label{fig:ROC_GN2_NDIVE}
\end{figure}

We set the constant $f_c$ to 0.05, which signifies the relative weight the discriminant will place on rejecting $c$ jets versus rejecting light-jets. This value can be tuned to give more or less weight to rejection of $c$- or light-jets, and is currently set to roughly the relative fraction of $c$-jets out of all background jets. The resulting efficiencies and rejection rates are shown in Fig.~\ref{fig:ROC_GN2_NDIVE}. Including \ndive\ in the flavour tagging model results in up to 40\% increased rejection rates of light-jets and up to 15\% increased rejection rates for $c$-jets. 

\begin{figure}[ht!]
\centering
    \begin{subfigure}[b]{0.4\textwidth}
    \centering
    \includegraphics[width=\textwidth]{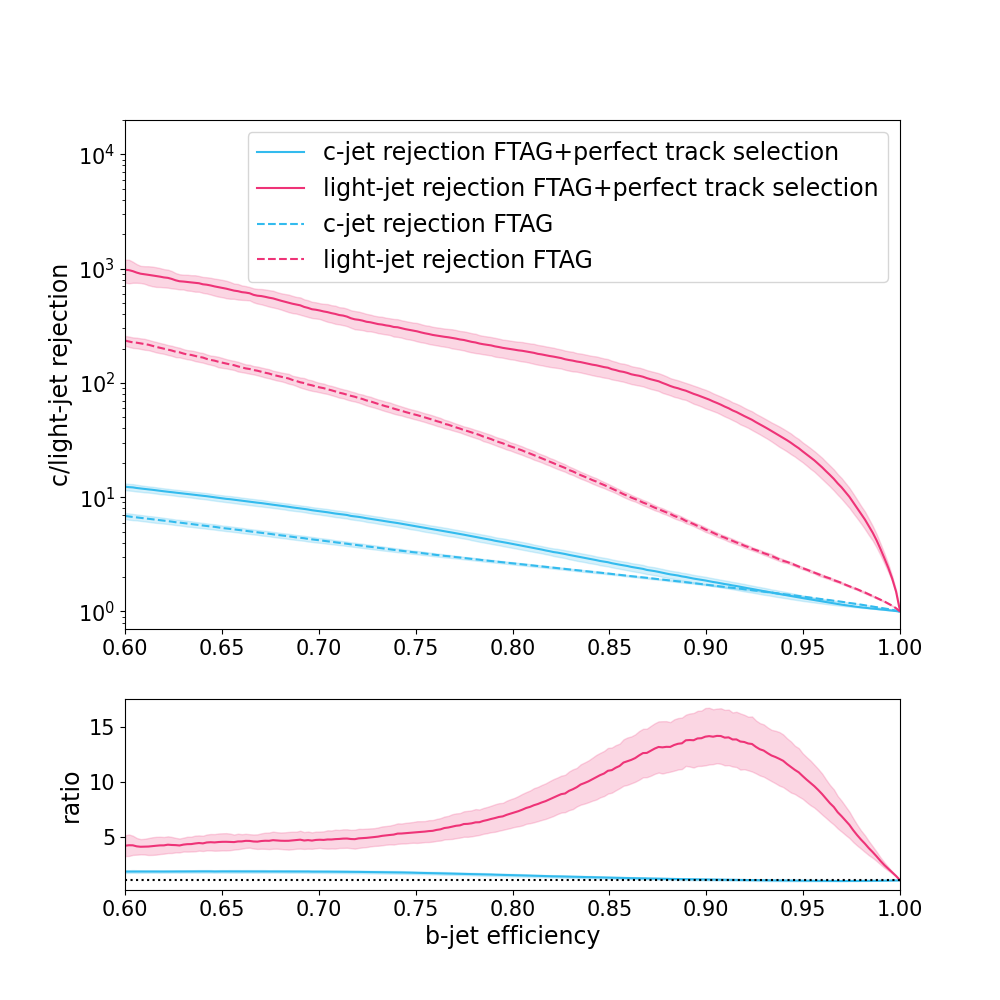}
    \end{subfigure}
\caption{Light-jet and $c$-jet rejection as a function of $b$-tagging efficiency for the flavour tagging models in the baseline architecture and with \ndive\ integration using a perfect track selection assignment. The error bars are the standard deviation of 5 trainings.} 
\label{fig:ROC_perfect}
\end{figure}

While flavour tagging performance gains are observed, we note that significant future performance improvements are possible with an \fndive\ like architecture. To see this, we show the potential for higher rejection rates to be achieved in the ideal scenario of the perfect track selection, where only the tracks originating from the secondary vertex are given to the vertex fit. The results in Fig.~\ref{fig:ROC_perfect} illustrate the potential for significant gains in rejection rates to be achieved with better track selection models.

\section{Conclusion}\label{sec:conclusion}
In this work, we introduce a new strategy to integrate vertex fitting algorithms into neural networks with differentiable programming. To do so, we make use of methods in differentiable optimization to enable differentiation through the optimization that defines the vertex fit.  

Using these techniques, we introduce the differentiable vertex fitting algorithm \ndive\ that is capable of finding and fitting secondary vertices in both $b$- and $c$-jets and can readily be integrated and jointly optimized in a larger flavour tagging neural network model. We have developed one possible flavour tagging model that integrates vertex fitting, \fndive, and show that considerable improvements in light-jet and $c$-jet rejection are possible when vertex fitting is used in neural flavour tagging models.  

These methodological developments are generic, applicable to other vertex fitting algorithms and other schemes for integrating vertex information into neural networks. The ultimate quality of a vertex-integrated model will depend on the vertex finding and fitting quality, and how the information is integrated into a larger flavour tagging model. As such, the models presented in this work represent one choice of how to use differentiable vertex fitting in neural flavour tagging models, but also show that large future improvements are possible. Increases in data set sizes, and consequently increases in transformer size for both the weight prediction and track processing modules, will also likely significantly improve model performance and are important directions of future work. Ultimately, we believe that using the differentiable vertex fitting strategy described here can lead to a highly fruitful future of continued improvement to neural flavour tagging models.

\begin{acknowledgments} 
We thank Nicole Hartman, Lukas Heinrich, Francesco Di Bello, and Rafael Teixeira de Lima for their helpful discussions and feedback on the manuscript.

MK, RS, and JS are supported by the US Department of Energy (DOE) under grant DE-AC02-76SF00515. RS is also supported by the US National Science Foundation (NSF) Graduate Research Fellowship Program.  IO and RI are supported by the fellowship LCF/BQ/PI20/11760025 from La Caixa Foundation (ID 100010434) and by the European Union Horizon 2020 research and innovation program under the Marie Sk{\l}odowska-Curie grant agreement No 847648.
\end{acknowledgments}

\bibliography{bibliography}

\begin{thebibliography}{54}%
\makeatletter
\providecommand \@ifxundefined [1]{%
 \@ifx{#1\undefined}
}%
\providecommand \@ifnum [1]{%
 \ifnum #1\expandafter \@firstoftwo
 \else \expandafter \@secondoftwo
 \fi
}%
\providecommand \@ifx [1]{%
 \ifx #1\expandafter \@firstoftwo
 \else \expandafter \@secondoftwo
 \fi
}%
\providecommand \natexlab [1]{#1}%
\providecommand \enquote  [1]{``#1''}%
\providecommand \bibnamefont  [1]{#1}%
\providecommand \bibfnamefont [1]{#1}%
\providecommand \citenamefont [1]{#1}%
\providecommand \href@noop [0]{\@secondoftwo}%
\providecommand \href [0]{\begingroup \@sanitize@url \@href}%
\providecommand \@href[1]{\@@startlink{#1}\@@href}%
\providecommand \@@href[1]{\endgroup#1\@@endlink}%
\providecommand \@sanitize@url [0]{\catcode `\\12\catcode `\$12\catcode
  `\&12\catcode `\#12\catcode `\^12\catcode `\_12\catcode `\%12\relax}%
\providecommand \@@startlink[1]{}%
\providecommand \@@endlink[0]{}%
\providecommand \url  [0]{\begingroup\@sanitize@url \@url }%
\providecommand \@url [1]{\endgroup\@href {#1}{\urlprefix }}%
\providecommand \urlprefix  [0]{URL }%
\providecommand \Eprint [0]{\href }%
\providecommand \doibase [0]{https://doi.org/}%
\providecommand \selectlanguage [0]{\@gobble}%
\providecommand \bibinfo  [0]{\@secondoftwo}%
\providecommand \bibfield  [0]{\@secondoftwo}%
\providecommand \translation [1]{[#1]}%
\providecommand \BibitemOpen [0]{}%
\providecommand \bibitemStop [0]{}%
\providecommand \bibitemNoStop [0]{.\EOS\space}%
\providecommand \EOS [0]{\spacefactor3000\relax}%
\providecommand \BibitemShut  [1]{\csname bibitem#1\endcsname}%
\let\auto@bib@innerbib\@empty
%</preamble>
\bibitem [{\citenamefont {Evans}\ and\ \citenamefont
  {Bryant}(2008)}]{Evans_2008}%
  \BibitemOpen
  \bibfield  {author} {\bibinfo {author} {\bibfnamefont {L.}~\bibnamefont
  {Evans}}\ and\ \bibinfo {author} {\bibfnamefont {P.}~\bibnamefont {Bryant}},\
  }\bibfield  {title} {\bibinfo {title} {{LHC} machine},\ }\href
  {https://doi.org/10.1088/1748-0221/3/08/S08001} {\bibfield  {journal}
  {\bibinfo  {journal} {Journal of Instrumentation}\ }\textbf {\bibinfo
  {volume} {3}}\bibinfo  {number} { (08)},\ \bibinfo {pages}
  {S08001}}\BibitemShut {NoStop}%
\bibitem [{\citenamefont {{ATLAS Collaboration}}(2016)}]{atlasbtag_2016}%
  \BibitemOpen
\bibfield  {number} {  }\bibfield  {author} {\bibinfo {author} {\bibnamefont
  {{ATLAS Collaboration}}},\ }\bibfield  {title} {\bibinfo {title}
  {{Performance of b-jet identification in the ATLAS experiment}},\ }\href
  {https://doi.org/10.1088/1748-0221/11/04/P04008} {\bibfield  {journal}
  {\bibinfo  {journal} {JINST}\ }\textbf {\bibinfo {volume} {11}}\bibinfo
  {number} { (04)},\ \bibinfo {pages} {P04008}}\BibitemShut {NoStop}%
\bibitem [{\citenamefont {{CMS collaboration}}(2013)}]{CMSbtag_2013}%
  \BibitemOpen
\bibfield  {number} {  }\bibfield  {author} {\bibinfo {author} {\bibnamefont
  {{CMS collaboration}}},\ }\bibfield  {title} {\bibinfo {title}
  {{Identification of b-quark jets with the CMS experiment}},\ }\href
  {https://doi.org/10.1088/1748-0221/8/04/P04013} {\bibfield  {journal}
  {\bibinfo  {journal} {JINST}\ }\textbf {\bibinfo {volume} {8}}\bibinfo
  {number} { (04)},\ \bibinfo {pages} {P04013}}\BibitemShut {NoStop}%
\bibitem [{\citenamefont {{CMS
  Collaboration}}(2018{\natexlab{a}})}]{Sirunyan_2018}%
  \BibitemOpen
\bibfield  {number} {  }\bibfield  {author} {\bibinfo {author} {\bibnamefont
  {{CMS Collaboration}}},\ }\bibfield  {title} {\bibinfo {title}
  {{Identification of heavy-flavour jets with the CMS detector in pp collisions
  at 13 TeV}},\ }\href {https://doi.org/10.1088/1748-0221/13/05/P05011}
  {\bibfield  {journal} {\bibinfo  {journal} {JINST}\ }\textbf {\bibinfo
  {volume} {13}}\bibinfo  {number} { (05)},\ \bibinfo {pages}
  {P05011}}\BibitemShut {NoStop}%
\bibitem [{\citenamefont {{ATLAS Collaboration}}(2019)}]{atlasbtag_2019}%
  \BibitemOpen
\bibfield  {number} {  }\bibfield  {author} {\bibinfo {author} {\bibnamefont
  {{ATLAS Collaboration}}},\ }\bibfield  {title} {\bibinfo {title} {Atlas b-jet
  identification performance and efficiency measurement with
  {\$}{\$}t{\{}{$\backslash$}bar{\{}t{\}}{\}}{\$}{\$} events in pp collisions
  at {\$}{\$}{$\backslash$}sqrt{\{}s{\}}=13{\$}{\$} tev},\ }\href@noop {}
  {\bibfield  {journal} {\bibinfo  {journal} {The European Physical Journal C}\
  }\textbf {\bibinfo {volume} {79}},\ \bibinfo {pages} {970} (\bibinfo {year}
  {2019})}\BibitemShut {NoStop}%
\bibitem [{\citenamefont {{ATLAS
  Collaboration}}(2023{\natexlab{a}})}]{atlasbtag_2023}%
  \BibitemOpen
  \bibfield  {author} {\bibinfo {author} {\bibnamefont {{ATLAS
  Collaboration}}},\ }\bibfield  {title} {\bibinfo {title} {Atlas
  flavour-tagging algorithms for the lhc run 2 pp collision dataset},\
  }\href@noop {} {\bibfield  {journal} {\bibinfo  {journal} {The European
  Physical Journal C}\ }\textbf {\bibinfo {volume} {83}},\ \bibinfo {pages}
  {681} (\bibinfo {year} {2023}{\natexlab{a}})}\BibitemShut {NoStop}%
\bibitem [{\citenamefont {{ATLAS
  Collaboration}}(2017)}]{ATL-PHYS-PUB-2017-003}%
  \BibitemOpen
  \bibfield  {author} {\bibinfo {author} {\bibnamefont {{ATLAS
  Collaboration}}},\ }\href@noop {} {\bibinfo {title} {{Identification of Jets
  Containing $b$-Hadrons with Recurrent Neural Networks at the ATLAS
  Experiment}}},\ \bibinfo {howpublished}
  {{\href{https://cds.cern.ch/record/2255226}{ATL-PHYS-PUB-2017-003}}}
  (\bibinfo {year} {2017})\BibitemShut {NoStop}%
\bibitem [{\citenamefont {Guest}\ \emph {et~al.}(2016)\citenamefont {Guest},
  \citenamefont {Collado}, \citenamefont {Baldi}, \citenamefont {Hsu},
  \citenamefont {Urban},\ and\ \citenamefont {Whiteson}}]{PhysRevD.94.112002}%
  \BibitemOpen
  \bibfield  {author} {\bibinfo {author} {\bibfnamefont {D.}~\bibnamefont
  {Guest}}, \bibinfo {author} {\bibfnamefont {J.}~\bibnamefont {Collado}},
  \bibinfo {author} {\bibfnamefont {P.}~\bibnamefont {Baldi}}, \bibinfo
  {author} {\bibfnamefont {S.-C.}\ \bibnamefont {Hsu}}, \bibinfo {author}
  {\bibfnamefont {G.}~\bibnamefont {Urban}},\ and\ \bibinfo {author}
  {\bibfnamefont {D.}~\bibnamefont {Whiteson}},\ }\bibfield  {title} {\bibinfo
  {title} {{Jet flavor classification in high-energy physics with deep neural
  networks}},\ }\href {https://doi.org/10.1103/PhysRevD.94.112002} {\bibfield
  {journal} {\bibinfo  {journal} {Phys. Rev. D}\ }\textbf {\bibinfo {volume}
  {94}},\ \bibinfo {pages} {112002} (\bibinfo {year} {2016})}\BibitemShut
  {NoStop}%
\bibitem [{\citenamefont {{CMS
  Collaboration}}(2018{\natexlab{b}})}]{CMS-DP-2018-058}%
  \BibitemOpen
  \bibfield  {author} {\bibinfo {author} {\bibnamefont {{CMS Collaboration}}},\
  }\href@noop {} {\bibinfo {title} {{Performance of the DeepJet b tagging
  algorithm using 41.9/fb of data from proton-proton collisions at 13TeV with
  Phase 1 CMS detector}}},\ \bibinfo {howpublished}
  {{\href{https://cds.cern.ch/record/2646773}{CMS-DP-2018-058}}} (\bibinfo
  {year} {2018}{\natexlab{b}})\BibitemShut {NoStop}%
\bibitem [{\citenamefont {Bols}\ \emph {et~al.}(2020)\citenamefont {Bols},
  \citenamefont {Kieseler}, \citenamefont {Verzetti}, \citenamefont {Stoye},\
  and\ \citenamefont {Stakia}}]{Bols_2020}%
  \BibitemOpen
  \bibfield  {author} {\bibinfo {author} {\bibfnamefont {E.}~\bibnamefont
  {Bols}}, \bibinfo {author} {\bibfnamefont {J.}~\bibnamefont {Kieseler}},
  \bibinfo {author} {\bibfnamefont {M.}~\bibnamefont {Verzetti}}, \bibinfo
  {author} {\bibfnamefont {M.}~\bibnamefont {Stoye}},\ and\ \bibinfo {author}
  {\bibfnamefont {A.}~\bibnamefont {Stakia}},\ }\bibfield  {title} {\bibinfo
  {title} {{Jet flavour classification using DeepJet}},\ }\href
  {https://doi.org/10.1088/1748-0221/15/12/P12012} {\bibfield  {journal}
  {\bibinfo  {journal} {JINST}\ }\textbf {\bibinfo {volume} {15}}\bibinfo
  {number} { (12)},\ \bibinfo {pages} {P12012}}\BibitemShut {NoStop}%
\bibitem [{\citenamefont {{ATLAS
  Collaboration}}(2020)}]{ATL-PHYS-PUB-2020-014}%
  \BibitemOpen
\bibfield  {number} {  }\bibfield  {author} {\bibinfo {author} {\bibnamefont
  {{ATLAS Collaboration}}},\ }\href@noop {} {\bibinfo {title} {{Deep Sets based
  Neural Networks for Impact Parameter Flavour Tagging in ATLAS}}},\ \bibinfo
  {howpublished}
  {{\href{https://cds.cern.ch/record/2718948}{ATL-PHYS-PUB-2020-014}}}
  (\bibinfo {year} {2020})\BibitemShut {NoStop}%
\bibitem [{\citenamefont {{ATLAS
  Collaboration}}(2022)}]{ATL-PHYS-PUB-2022-027}%
  \BibitemOpen
  \bibfield  {author} {\bibinfo {author} {\bibnamefont {{ATLAS
  Collaboration}}},\ }\href@noop {} {\bibinfo {title} {{Graph Neural Network
  Jet Flavour Tagging with the ATLAS Detector}}},\ \bibinfo {howpublished}
  {{\href{https://cds.cern.ch/record/2811135}{ATL-PHYS-PUB-2022-027}}}
  (\bibinfo {year} {2022})\BibitemShut {NoStop}%
\bibitem [{\citenamefont {{ATLAS
  Collaboration}}(2023{\natexlab{b}})}]{ATL-PLOT-FTAG-2023-01}%
  \BibitemOpen
  \bibfield  {author} {\bibinfo {author} {\bibnamefont {{ATLAS
  Collaboration}}},\ }\href@noop {} {\bibinfo {title} {{Jet Flavour Tagging
  With GN1 and DL1d. Generator dependence, Run 2 and Run 3 data agreement
  studies}}},\ \bibinfo {howpublished}
  {{\href{https://atlas.web.cern.ch/Atlas/GROUPS/PHYSICS/PLOTS/FTAG-2023-01/}{ATL-PLOT-FTAG-2023-01}}}
  (\bibinfo {year} {2023}{\natexlab{b}})\BibitemShut {NoStop}%
\bibitem [{\citenamefont {{ATLAS
  Collaboration}}(2023{\natexlab{c}})}]{ATL-PHYS-PUB-2023-021}%
  \BibitemOpen
  \bibfield  {author} {\bibinfo {author} {\bibnamefont {{ATLAS
  Collaboration}}},\ }\href@noop {} {\bibinfo {title} {{Transformer Neural
  Networks for Identifying Boosted Higgs Bosons decaying into $b\bar{b}$ and
  $c\bar{c}$ in ATLAS}}},\ \bibinfo {howpublished}
  {{\href{https://cds.cern.ch/record/2866601}{ATL-PHYS-PUB-2023-021}}}
  (\bibinfo {year} {2023}{\natexlab{c}})\BibitemShut {NoStop}%
\bibitem [{\citenamefont {{CMS Collaboration}}(2020)}]{CMS-DP-2020-002}%
  \BibitemOpen
  \bibfield  {author} {\bibinfo {author} {\bibnamefont {{CMS Collaboration}}},\
  }\href@noop {} {\bibinfo {title} {{Identification of highly Lorentz-boosted
  heavy particles using graph neural networks and new mass decorrelation
  techniques}}},\ \bibinfo {howpublished}
  {{\href{https://cds.cern.ch/record/2707946}{CMS-DP-2020-002}}} (\bibinfo
  {year} {2020})\BibitemShut {NoStop}%
\bibitem [{\citenamefont {Moreno}\ \emph {et~al.}(2020)\citenamefont {Moreno},
  \citenamefont {Nguyen}, \citenamefont {Vlimant}, \citenamefont {Cerri},
  \citenamefont {Newman}, \citenamefont {Periwal}, \citenamefont {Spiropulu},
  \citenamefont {Duarte},\ and\ \citenamefont {Pierini}}]{PhysRevD.102.012010}%
  \BibitemOpen
  \bibfield  {author} {\bibinfo {author} {\bibfnamefont {E.~A.}\ \bibnamefont
  {Moreno}}, \bibinfo {author} {\bibfnamefont {T.~Q.}\ \bibnamefont {Nguyen}},
  \bibinfo {author} {\bibfnamefont {J.-R.}\ \bibnamefont {Vlimant}}, \bibinfo
  {author} {\bibfnamefont {O.}~\bibnamefont {Cerri}}, \bibinfo {author}
  {\bibfnamefont {H.~B.}\ \bibnamefont {Newman}}, \bibinfo {author}
  {\bibfnamefont {A.}~\bibnamefont {Periwal}}, \bibinfo {author} {\bibfnamefont
  {M.}~\bibnamefont {Spiropulu}}, \bibinfo {author} {\bibfnamefont {J.~M.}\
  \bibnamefont {Duarte}},\ and\ \bibinfo {author} {\bibfnamefont
  {M.}~\bibnamefont {Pierini}},\ }\bibfield  {title} {\bibinfo {title}
  {Interaction networks for the identification of boosted
  $h\ensuremath{\rightarrow}b\overline{b}$ decays},\ }\href
  {https://doi.org/10.1103/PhysRevD.102.012010} {\bibfield  {journal} {\bibinfo
   {journal} {Phys. Rev. D}\ }\textbf {\bibinfo {volume} {102}},\ \bibinfo
  {pages} {012010} (\bibinfo {year} {2020})}\BibitemShut {NoStop}%
\bibitem [{\citenamefont {Shlomi}\ \emph {et~al.}(2021)\citenamefont {Shlomi},
  \citenamefont {Ganguly}, \citenamefont {Gross}, \citenamefont {Cranmer},
  \citenamefont {Lipman}, \citenamefont {Serviansky}, \citenamefont {Maron},\
  and\ \citenamefont {Segol}}]{Shlomi_2021}%
  \BibitemOpen
  \bibfield  {author} {\bibinfo {author} {\bibfnamefont {J.}~\bibnamefont
  {Shlomi}}, \bibinfo {author} {\bibfnamefont {S.}~\bibnamefont {Ganguly}},
  \bibinfo {author} {\bibfnamefont {E.}~\bibnamefont {Gross}}, \bibinfo
  {author} {\bibfnamefont {K.}~\bibnamefont {Cranmer}}, \bibinfo {author}
  {\bibfnamefont {Y.}~\bibnamefont {Lipman}}, \bibinfo {author} {\bibfnamefont
  {H.}~\bibnamefont {Serviansky}}, \bibinfo {author} {\bibfnamefont
  {H.}~\bibnamefont {Maron}},\ and\ \bibinfo {author} {\bibfnamefont
  {N.}~\bibnamefont {Segol}},\ }\bibfield  {title} {\bibinfo {title}
  {{Secondary vertex finding in jets with neural networks}},\ }\href
  {https://doi.org/10.1140/epjc/s10052-021-09342-y} {\bibfield  {journal}
  {\bibinfo  {journal} {EPJC}\ }\textbf {\bibinfo {volume} {81}},\ \bibinfo
  {pages} {540} (\bibinfo {year} {2021})}\BibitemShut {NoStop}%
\bibitem [{\citenamefont {Goto}\ \emph {et~al.}(2023)\citenamefont {Goto},
  \citenamefont {Suehara}, \citenamefont {Yoshioka}, \citenamefont {Kurata},
  \citenamefont {Nagahara}, \citenamefont {Nakashima}, \citenamefont
  {Takemura},\ and\ \citenamefont {Iwasaki}}]{GOTO2023167836}%
  \BibitemOpen
  \bibfield  {author} {\bibinfo {author} {\bibfnamefont {K.}~\bibnamefont
  {Goto}}, \bibinfo {author} {\bibfnamefont {T.}~\bibnamefont {Suehara}},
  \bibinfo {author} {\bibfnamefont {T.}~\bibnamefont {Yoshioka}}, \bibinfo
  {author} {\bibfnamefont {M.}~\bibnamefont {Kurata}}, \bibinfo {author}
  {\bibfnamefont {H.}~\bibnamefont {Nagahara}}, \bibinfo {author}
  {\bibfnamefont {Y.}~\bibnamefont {Nakashima}}, \bibinfo {author}
  {\bibfnamefont {N.}~\bibnamefont {Takemura}},\ and\ \bibinfo {author}
  {\bibfnamefont {M.}~\bibnamefont {Iwasaki}},\ }\bibfield  {title} {\bibinfo
  {title} {{Development of a vertex finding algorithm using Recurrent Neural
  Network}},\ }\href
  {https://doi.org/https://doi.org/10.1016/j.nima.2022.167836} {\bibfield
  {journal} {\bibinfo  {journal} {NIM A}\ }\textbf {\bibinfo {volume} {1047}},\
  \bibinfo {pages} {167836} (\bibinfo {year} {2023})}\BibitemShut {NoStop}%
\bibitem [{\citenamefont {Bronstein}\ \emph {et~al.}(2021)\citenamefont
  {Bronstein}, \citenamefont {Bruna}, \citenamefont {Cohen},\ and\
  \citenamefont {Velickovic}}]{DBLP:journals/corr/abs-2104-13478}%
  \BibitemOpen
  \bibfield  {author} {\bibinfo {author} {\bibfnamefont {M.~M.}\ \bibnamefont
  {Bronstein}}, \bibinfo {author} {\bibfnamefont {J.}~\bibnamefont {Bruna}},
  \bibinfo {author} {\bibfnamefont {T.}~\bibnamefont {Cohen}},\ and\ \bibinfo
  {author} {\bibfnamefont {P.}~\bibnamefont {Velickovic}},\ }\bibfield  {title}
  {\bibinfo {title} {Geometric deep learning: Grids, groups, graphs, geodesics,
  and gauges},\ }\href {https://arxiv.org/abs/2104.13478} {\bibfield  {journal}
  {\bibinfo  {journal} {CoRR}\ }\textbf {\bibinfo {volume} {abs/2104.13478}}
  (\bibinfo {year} {2021})},\ \Eprint {https://arxiv.org/abs/2104.13478}
  {2104.13478} \BibitemShut {NoStop}%
\bibitem [{\citenamefont {Vaswani}\ \emph {et~al.}(2017)\citenamefont
  {Vaswani}, \citenamefont {Shazeer}, \citenamefont {Parmar}, \citenamefont
  {Uszkoreit}, \citenamefont {Jones}, \citenamefont {Gomez}, \citenamefont
  {Kaiser},\ and\ \citenamefont {Polosukhin}}]{NIPS2017_3f5ee243}%
  \BibitemOpen
  \bibfield  {author} {\bibinfo {author} {\bibfnamefont {A.}~\bibnamefont
  {Vaswani}}, \bibinfo {author} {\bibfnamefont {N.}~\bibnamefont {Shazeer}},
  \bibinfo {author} {\bibfnamefont {N.}~\bibnamefont {Parmar}}, \bibinfo
  {author} {\bibfnamefont {J.}~\bibnamefont {Uszkoreit}}, \bibinfo {author}
  {\bibfnamefont {L.}~\bibnamefont {Jones}}, \bibinfo {author} {\bibfnamefont
  {A.~N.}\ \bibnamefont {Gomez}}, \bibinfo {author} {\bibfnamefont {L.~u.}\
  \bibnamefont {Kaiser}},\ and\ \bibinfo {author} {\bibfnamefont
  {I.}~\bibnamefont {Polosukhin}},\ }\bibfield  {title} {\bibinfo {title}
  {Attention is all you need},\ }in\ \href
  {https://proceedings.neurips.cc/paper_files/paper/2017/file/3f5ee243547dee91fbd053c1c4a845aa-Paper.pdf}
  {\emph {\bibinfo {booktitle} {Advances in Neural Information Processing
  Systems}}},\ Vol.~\bibinfo {volume} {30},\ \bibinfo {editor} {edited by\
  \bibinfo {editor} {\bibfnamefont {I.}~\bibnamefont {Guyon}}, \bibinfo
  {editor} {\bibfnamefont {U.~V.}\ \bibnamefont {Luxburg}}, \bibinfo {editor}
  {\bibfnamefont {S.}~\bibnamefont {Bengio}}, \bibinfo {editor} {\bibfnamefont
  {H.}~\bibnamefont {Wallach}}, \bibinfo {editor} {\bibfnamefont
  {R.}~\bibnamefont {Fergus}}, \bibinfo {editor} {\bibfnamefont
  {S.}~\bibnamefont {Vishwanathan}},\ and\ \bibinfo {editor} {\bibfnamefont
  {R.}~\bibnamefont {Garnett}}}\ (\bibinfo  {publisher} {Curran Associates,
  Inc.},\ \bibinfo {year} {2017})\BibitemShut {NoStop}%
\bibitem [{\citenamefont {Billoir}\ \emph {et~al.}(1985)\citenamefont
  {Billoir}, \citenamefont {Fr{\"y}hwirth},\ and\ \citenamefont
  {Regler}}]{BILLOIR1985115}%
  \BibitemOpen
  \bibfield  {author} {\bibinfo {author} {\bibfnamefont {P.}~\bibnamefont
  {Billoir}}, \bibinfo {author} {\bibfnamefont {R.}~\bibnamefont
  {Fr{\"y}hwirth}},\ and\ \bibinfo {author} {\bibfnamefont {M.}~\bibnamefont
  {Regler}},\ }\bibfield  {title} {\bibinfo {title} {{Track element merging
  strategy and vertex fitting in complex modular detectors}},\ }\href
  {https://doi.org/https://doi.org/10.1016/0168-9002(85)90523-6} {\bibfield
  {journal} {\bibinfo  {journal} {NIM A}\ }\textbf {\bibinfo {volume} {241}},\
  \bibinfo {pages} {115} (\bibinfo {year} {1985})}\BibitemShut {NoStop}%
\bibitem [{\citenamefont {Fruhwirth}\ \emph {et~al.}(2007)\citenamefont
  {Fruhwirth}, \citenamefont {Waltenberger},\ and\ \citenamefont
  {Vanlaer}}]{Fruhwirth:2007hz}%
  \BibitemOpen
  \bibfield  {author} {\bibinfo {author} {\bibfnamefont {R.}~\bibnamefont
  {Fruhwirth}}, \bibinfo {author} {\bibfnamefont {W.}~\bibnamefont
  {Waltenberger}},\ and\ \bibinfo {author} {\bibfnamefont {P.}~\bibnamefont
  {Vanlaer}},\ }\bibfield  {title} {\bibinfo {title} {{Adaptive vertex
  fitting}},\ }\href {https://doi.org/10.1088/0954-3899/34/12/N01} {\bibfield
  {journal} {\bibinfo  {journal} {J. Phys. G}\ }\textbf {\bibinfo {volume}
  {34}},\ \bibinfo {pages} {N343} (\bibinfo {year} {2007})}\BibitemShut
  {NoStop}%
\bibitem [{\citenamefont {Fr{\"u}wirth}\ and\ \citenamefont
  {Strandlie}(2021)}]{Fruhwirth2021}%
  \BibitemOpen
  \bibfield  {author} {\bibinfo {author} {\bibfnamefont {R.}~\bibnamefont
  {Fr{\"u}wirth}}\ and\ \bibinfo {author} {\bibfnamefont {A.}~\bibnamefont
  {Strandlie}},\ }\href
  {https://doi.org/https://doi.org/10.1007/978-3-030-65771-0} {\emph {\bibinfo
  {title} {{Pattern Recognition, Tracking and Vertex Reconstruction in Particle
  Detectors}}}}\ (\bibinfo  {publisher} {{Springer}},\ \bibinfo {year}
  {2021})\BibitemShut {NoStop}%
\bibitem [{\citenamefont {Piacquadio}(2010)}]{Giacinto2010}%
  \BibitemOpen
  \bibfield  {author} {\bibinfo {author} {\bibfnamefont {G.}~\bibnamefont
  {Piacquadio}},\ }\emph {\bibinfo {title} {{Identification of b-jets and
  investigation of the discovery potential of a Higgs boson in the $WH\to l\nu
  b\bar{b}$ channel with the ATLAS experiment}}},\ \href
  {https://cds.cern.ch/record/1243771} {Ph.D. thesis},\ \bibinfo  {school}
  {Freiburg University} (\bibinfo {year} {2010})\BibitemShut {NoStop}%
\bibitem [{\citenamefont {Bortolotto}\ \emph {et~al.}(1991)\citenamefont
  {Bortolotto}, \citenamefont {{De Angelis}},\ and\ \citenamefont
  {Lanceri}}]{BORTOLOTTO1991459}%
  \BibitemOpen
  \bibfield  {author} {\bibinfo {author} {\bibfnamefont {C.}~\bibnamefont
  {Bortolotto}}, \bibinfo {author} {\bibfnamefont {A.}~\bibnamefont {{De
  Angelis}}},\ and\ \bibinfo {author} {\bibfnamefont {L.}~\bibnamefont
  {Lanceri}},\ }\bibfield  {title} {\bibinfo {title} {{Tagging the decays of
  the Z0 boson into b quark pairs with a neural network classifier}},\ }\href
  {https://doi.org/https://doi.org/10.1016/0168-9002(91)90039-S} {\bibfield
  {journal} {\bibinfo  {journal} {NIM A}\ }\textbf {\bibinfo {volume} {306}},\
  \bibinfo {pages} {459} (\bibinfo {year} {1991})}\BibitemShut {NoStop}%
\bibitem [{\citenamefont {Qu}\ and\ \citenamefont
  {Gouskos}(2020)}]{PhysRevD.101.056019}%
  \BibitemOpen
  \bibfield  {author} {\bibinfo {author} {\bibfnamefont {H.}~\bibnamefont
  {Qu}}\ and\ \bibinfo {author} {\bibfnamefont {L.}~\bibnamefont {Gouskos}},\
  }\bibfield  {title} {\bibinfo {title} {Jet tagging via particle clouds},\
  }\href {https://doi.org/10.1103/PhysRevD.101.056019} {\bibfield  {journal}
  {\bibinfo  {journal} {Phys. Rev. D}\ }\textbf {\bibinfo {volume} {101}},\
  \bibinfo {pages} {056019} (\bibinfo {year} {2020})}\BibitemShut {NoStop}%
\bibitem [{\citenamefont {Qu}\ \emph {et~al.}(2022)\citenamefont {Qu},
  \citenamefont {Li},\ and\ \citenamefont {Qian}}]{DBLP:conf/icml/QuLQ22}%
  \BibitemOpen
  \bibfield  {author} {\bibinfo {author} {\bibfnamefont {H.}~\bibnamefont
  {Qu}}, \bibinfo {author} {\bibfnamefont {C.}~\bibnamefont {Li}},\ and\
  \bibinfo {author} {\bibfnamefont {S.}~\bibnamefont {Qian}},\ }\bibfield
  {title} {\bibinfo {title} {Particle transformer for jet tagging},\ }in\ \href
  {https://proceedings.mlr.press/v162/qu22b.html} {\emph {\bibinfo {booktitle}
  {International Conference on Machine Learning, {ICML} 2022, 17-23 July 2022,
  Baltimore, Maryland, {USA}}}},\ \bibinfo {series} {Proceedings of Machine
  Learning Research}, Vol.\ \bibinfo {volume} {162},\ \bibinfo {editor} {edited
  by\ \bibinfo {editor} {\bibfnamefont {K.}~\bibnamefont {Chaudhuri}}, \bibinfo
  {editor} {\bibfnamefont {S.}~\bibnamefont {Jegelka}}, \bibinfo {editor}
  {\bibfnamefont {L.}~\bibnamefont {Song}}, \bibinfo {editor} {\bibfnamefont
  {C.}~\bibnamefont {Szepesv{\'{a}}ri}}, \bibinfo {editor} {\bibfnamefont
  {G.}~\bibnamefont {Niu}},\ and\ \bibinfo {editor} {\bibfnamefont
  {S.}~\bibnamefont {Sabato}}}\ (\bibinfo  {publisher} {{PMLR}},\ \bibinfo
  {year} {2022})\ pp.\ \bibinfo {pages} {18281--18292}\BibitemShut {NoStop}%
\bibitem [{\citenamefont {Amos}\ and\ \citenamefont
  {Kolter}(2017)}]{10.5555/3305381.3305396}%
  \BibitemOpen
  \bibfield  {author} {\bibinfo {author} {\bibfnamefont {B.}~\bibnamefont
  {Amos}}\ and\ \bibinfo {author} {\bibfnamefont {J.~Z.}\ \bibnamefont
  {Kolter}},\ }\bibfield  {title} {\bibinfo {title} {Optnet: Differentiable
  optimization as a layer in neural networks},\ }in\ \href@noop {} {\emph
  {\bibinfo {booktitle} {ICML}}},\ \bibinfo {series and number} {ICML'17}\
  (\bibinfo  {publisher} {JMLR.org},\ \bibinfo {year} {2017})\ p.\ \bibinfo
  {pages} {136–145}\BibitemShut {NoStop}%
\bibitem [{\citenamefont {Agrawal}\ \emph {et~al.}(2019)\citenamefont
  {Agrawal}, \citenamefont {Amos}, \citenamefont {Barratt}, \citenamefont
  {Boyd}, \citenamefont {Diamond},\ and\ \citenamefont
  {Kolter}}]{NEURIPS2019_9ce3c52f}%
  \BibitemOpen
  \bibfield  {author} {\bibinfo {author} {\bibfnamefont {A.}~\bibnamefont
  {Agrawal}}, \bibinfo {author} {\bibfnamefont {B.}~\bibnamefont {Amos}},
  \bibinfo {author} {\bibfnamefont {S.}~\bibnamefont {Barratt}}, \bibinfo
  {author} {\bibfnamefont {S.}~\bibnamefont {Boyd}}, \bibinfo {author}
  {\bibfnamefont {S.}~\bibnamefont {Diamond}},\ and\ \bibinfo {author}
  {\bibfnamefont {J.~Z.}\ \bibnamefont {Kolter}},\ }\bibfield  {title}
  {\bibinfo {title} {Differentiable convex optimization layers},\ }in\ \href
  {https://proceedings.neurips.cc/paper_files/paper/2019/file/9ce3c52fc54362e22053399d3181c638-Paper.pdf}
  {\emph {\bibinfo {booktitle} {NeurIPS}}},\ Vol.~\bibinfo {volume} {32},\
  \bibinfo {editor} {edited by\ \bibinfo {editor} {\bibfnamefont
  {H.}~\bibnamefont {Wallach}}, \bibinfo {editor} {\bibfnamefont
  {H.}~\bibnamefont {Larochelle}}, \bibinfo {editor} {\bibfnamefont
  {A.}~\bibnamefont {Beygelzimer}}, \bibinfo {editor} {\bibfnamefont
  {F.}~\bibnamefont {d\textquotesingle Alch\'{e}-Buc}}, \bibinfo {editor}
  {\bibfnamefont {E.}~\bibnamefont {Fox}},\ and\ \bibinfo {editor}
  {\bibfnamefont {R.}~\bibnamefont {Garnett}}}\ (\bibinfo  {publisher} {Curran
  Associates, Inc.},\ \bibinfo {year} {2019})\BibitemShut {NoStop}%
\bibitem [{\citenamefont {Diamond}\ and\ \citenamefont {Boyd}(2016)}]{cvxpy}%
  \BibitemOpen
  \bibfield  {author} {\bibinfo {author} {\bibfnamefont {S.}~\bibnamefont
  {Diamond}}\ and\ \bibinfo {author} {\bibfnamefont {S.}~\bibnamefont {Boyd}},\
  }\bibfield  {title} {\bibinfo {title} {{CVXPY}: A {P}ython-embedded modeling
  language for convex optimization},\ }\href
  {https://stanford.edu/~boyd/papers/pdf/cvxpy_paper.pdf} {\bibfield  {journal}
  {\bibinfo  {journal} {JMLR}\ } (\bibinfo {year} {2016})},\ \bibinfo {note}
  {to appear}\BibitemShut {NoStop}%
\bibitem [{\citenamefont {Kolter}\ \emph {et~al.}(2020)\citenamefont {Kolter},
  \citenamefont {Duvenaud},\ and\ \citenamefont {Johnson}}]{impl_tutorial}%
  \BibitemOpen
  \bibfield  {author} {\bibinfo {author} {\bibfnamefont {Z.}~\bibnamefont
  {Kolter}}, \bibinfo {author} {\bibfnamefont {D.}~\bibnamefont {Duvenaud}},\
  and\ \bibinfo {author} {\bibfnamefont {M.}~\bibnamefont {Johnson}},\ }\href
  {http://implicit-layers-tutorial.org/} {\bibinfo {title} {{Deep Implicit
  Layers - Neural ODEs, Deep Equilibirum Models, and Beyond}}},\ \bibinfo
  {howpublished} {NeurIPS 2020 tutorial} (\bibinfo {year} {2020})\BibitemShut
  {NoStop}%
\bibitem [{\citenamefont {Krantz}(2013)}]{Krantz2013}%
  \BibitemOpen
  \bibfield  {author} {\bibinfo {author} {\bibfnamefont {H.}~\bibnamefont
  {Krantz}, \bibfnamefont {Steven amd~Parks}},\ }\href
  {https://doi.org/https://doi.org/10.1007/978-1-4614-5981-1} {\emph {\bibinfo
  {title} {{The Implicit Function Theorem: History, Theory, and
  Applications}}}}\ (\bibinfo  {publisher} {{Birkh{\"a}user}},\ \bibinfo {year}
  {2013})\BibitemShut {NoStop}%
\bibitem [{\citenamefont {Blondel}\ \emph {et~al.}(2022)\citenamefont
  {Blondel}, \citenamefont {Berthet}, \citenamefont {Cuturi}, \citenamefont
  {Frostig}, \citenamefont {Hoyer}, \citenamefont {Llinares-Lopez},
  \citenamefont {Pedregosa},\ and\ \citenamefont
  {Vert}}]{NEURIPS2022_228b9279}%
  \BibitemOpen
  \bibfield  {author} {\bibinfo {author} {\bibfnamefont {M.}~\bibnamefont
  {Blondel}}, \bibinfo {author} {\bibfnamefont {Q.}~\bibnamefont {Berthet}},
  \bibinfo {author} {\bibfnamefont {M.}~\bibnamefont {Cuturi}}, \bibinfo
  {author} {\bibfnamefont {R.}~\bibnamefont {Frostig}}, \bibinfo {author}
  {\bibfnamefont {S.}~\bibnamefont {Hoyer}}, \bibinfo {author} {\bibfnamefont
  {F.}~\bibnamefont {Llinares-Lopez}}, \bibinfo {author} {\bibfnamefont
  {F.}~\bibnamefont {Pedregosa}},\ and\ \bibinfo {author} {\bibfnamefont
  {J.-P.}\ \bibnamefont {Vert}},\ }\bibfield  {title} {\bibinfo {title}
  {Efficient and modular implicit differentiation},\ }in\ \href
  {https://proceedings.neurips.cc/paper_files/paper/2022/file/228b9279ecf9bbafe582406850c57115-Paper-Conference.pdf}
  {\emph {\bibinfo {booktitle} {NeurIPS}}},\ Vol.~\bibinfo {volume} {35},\
  \bibinfo {editor} {edited by\ \bibinfo {editor} {\bibfnamefont
  {S.}~\bibnamefont {Koyejo}}, \bibinfo {editor} {\bibfnamefont
  {S.}~\bibnamefont {Mohamed}}, \bibinfo {editor} {\bibfnamefont
  {A.}~\bibnamefont {Agarwal}}, \bibinfo {editor} {\bibfnamefont
  {D.}~\bibnamefont {Belgrave}}, \bibinfo {editor} {\bibfnamefont
  {K.}~\bibnamefont {Cho}},\ and\ \bibinfo {editor} {\bibfnamefont
  {A.}~\bibnamefont {Oh}}}\ (\bibinfo  {publisher} {Curran Associates, Inc.},\
  \bibinfo {year} {2022})\ pp.\ \bibinfo {pages} {5230--5242}\BibitemShut
  {NoStop}%
\bibitem [{\citenamefont {B{\"u}cker}\ \emph {et~al.}(2005)\citenamefont
  {B{\"u}cker}, \citenamefont {Corliss}, \citenamefont {Hovland}, \citenamefont
  {Naumann},\ and\ \citenamefont {Norris}}]{autodiff}%
  \BibitemOpen
  \bibinfo {editor} {\bibfnamefont {H.~M.}\ \bibnamefont {B{\"u}cker}},
  \bibinfo {editor} {\bibfnamefont {G.~F.}\ \bibnamefont {Corliss}}, \bibinfo
  {editor} {\bibfnamefont {P.~D.}\ \bibnamefont {Hovland}}, \bibinfo {editor}
  {\bibfnamefont {U.}~\bibnamefont {Naumann}},\ and\ \bibinfo {editor}
  {\bibfnamefont {B.}~\bibnamefont {Norris}},\ eds.,\ \href
  {https://doi.org/https://doi.org/10.1007/3-540-28438-9} {\emph {\bibinfo
  {title} {Automatic Differentiation: {A}pplications, Theory, and
  Implementations}}},\ Lecture Notes in Computational Science and Engineering\
  (\bibinfo  {publisher} {Springer},\ \bibinfo {address} {New York, NY},\
  \bibinfo {year} {2005})\BibitemShut {NoStop}%
\bibitem [{\citenamefont {Baydin}\ \emph {et~al.}(2018)\citenamefont {Baydin},
  \citenamefont {Pearlmutter}, \citenamefont {Radul},\ and\ \citenamefont
  {Siskind}}]{JMLR:v18:17-468}%
  \BibitemOpen
  \bibfield  {author} {\bibinfo {author} {\bibfnamefont {A.~G.}\ \bibnamefont
  {Baydin}}, \bibinfo {author} {\bibfnamefont {B.~A.}\ \bibnamefont
  {Pearlmutter}}, \bibinfo {author} {\bibfnamefont {A.~A.}\ \bibnamefont
  {Radul}},\ and\ \bibinfo {author} {\bibfnamefont {J.~M.}\ \bibnamefont
  {Siskind}},\ }\bibfield  {title} {\bibinfo {title} {Automatic differentiation
  in machine learning: a survey},\ }\href
  {http://jmlr.org/papers/v18/17-468.html} {\bibfield  {journal} {\bibinfo
  {journal} {JMLR}\ }\textbf {\bibinfo {volume} {18}},\ \bibinfo {pages} {1}
  (\bibinfo {year} {2018})}\BibitemShut {NoStop}%
\bibitem [{\citenamefont {Abadi}\ \emph {et~al.}(2015)\citenamefont {Abadi},
  \citenamefont {Agarwal}, \citenamefont {Barham}, \citenamefont {Brevdo},
  \citenamefont {Chen}, \citenamefont {Citro}, \citenamefont {Corrado},
  \citenamefont {Davis}, \citenamefont {Dean}, \citenamefont {Devin},
  \citenamefont {Ghemawat}, \citenamefont {Goodfellow}, \citenamefont {Harp},
  \citenamefont {Irving}, \citenamefont {Isard}, \citenamefont {Jia},
  \citenamefont {Jozefowicz}, \citenamefont {Kaiser}, \citenamefont {Kudlur},
  \citenamefont {Levenberg}, \citenamefont {Man\'{e}}, \citenamefont {Monga},
  \citenamefont {Moore}, \citenamefont {Murray}, \citenamefont {Olah},
  \citenamefont {Schuster}, \citenamefont {Shlens}, \citenamefont {Steiner},
  \citenamefont {Sutskever}, \citenamefont {Talwar}, \citenamefont {Tucker},
  \citenamefont {Vanhoucke}, \citenamefont {Vasudevan}, \citenamefont
  {Vi\'{e}gas}, \citenamefont {Vinyals}, \citenamefont {Warden}, \citenamefont
  {Wattenberg}, \citenamefont {Wicke}, \citenamefont {Yu},\ and\ \citenamefont
  {Zheng}}]{tensorflow2015-whitepaper}%
  \BibitemOpen
  \bibfield  {author} {\bibinfo {author} {\bibfnamefont {M.}~\bibnamefont
  {Abadi}}, \bibinfo {author} {\bibfnamefont {A.}~\bibnamefont {Agarwal}},
  \bibinfo {author} {\bibfnamefont {P.}~\bibnamefont {Barham}}, \bibinfo
  {author} {\bibfnamefont {E.}~\bibnamefont {Brevdo}}, \bibinfo {author}
  {\bibfnamefont {Z.}~\bibnamefont {Chen}}, \bibinfo {author} {\bibfnamefont
  {C.}~\bibnamefont {Citro}}, \bibinfo {author} {\bibfnamefont {G.~S.}\
  \bibnamefont {Corrado}}, \bibinfo {author} {\bibfnamefont {A.}~\bibnamefont
  {Davis}}, \bibinfo {author} {\bibfnamefont {J.}~\bibnamefont {Dean}},
  \bibinfo {author} {\bibfnamefont {M.}~\bibnamefont {Devin}}, \bibinfo
  {author} {\bibfnamefont {S.}~\bibnamefont {Ghemawat}}, \bibinfo {author}
  {\bibfnamefont {I.}~\bibnamefont {Goodfellow}}, \bibinfo {author}
  {\bibfnamefont {A.}~\bibnamefont {Harp}}, \bibinfo {author} {\bibfnamefont
  {G.}~\bibnamefont {Irving}}, \bibinfo {author} {\bibfnamefont
  {M.}~\bibnamefont {Isard}}, \bibinfo {author} {\bibfnamefont
  {Y.}~\bibnamefont {Jia}}, \bibinfo {author} {\bibfnamefont {R.}~\bibnamefont
  {Jozefowicz}}, \bibinfo {author} {\bibfnamefont {L.}~\bibnamefont {Kaiser}},
  \bibinfo {author} {\bibfnamefont {M.}~\bibnamefont {Kudlur}}, \bibinfo
  {author} {\bibfnamefont {J.}~\bibnamefont {Levenberg}}, \bibinfo {author}
  {\bibfnamefont {D.}~\bibnamefont {Man\'{e}}}, \bibinfo {author}
  {\bibfnamefont {R.}~\bibnamefont {Monga}}, \bibinfo {author} {\bibfnamefont
  {S.}~\bibnamefont {Moore}}, \bibinfo {author} {\bibfnamefont
  {D.}~\bibnamefont {Murray}}, \bibinfo {author} {\bibfnamefont
  {C.}~\bibnamefont {Olah}}, \bibinfo {author} {\bibfnamefont {M.}~\bibnamefont
  {Schuster}}, \bibinfo {author} {\bibfnamefont {J.}~\bibnamefont {Shlens}},
  \bibinfo {author} {\bibfnamefont {B.}~\bibnamefont {Steiner}}, \bibinfo
  {author} {\bibfnamefont {I.}~\bibnamefont {Sutskever}}, \bibinfo {author}
  {\bibfnamefont {K.}~\bibnamefont {Talwar}}, \bibinfo {author} {\bibfnamefont
  {P.}~\bibnamefont {Tucker}}, \bibinfo {author} {\bibfnamefont
  {V.}~\bibnamefont {Vanhoucke}}, \bibinfo {author} {\bibfnamefont
  {V.}~\bibnamefont {Vasudevan}}, \bibinfo {author} {\bibfnamefont
  {F.}~\bibnamefont {Vi\'{e}gas}}, \bibinfo {author} {\bibfnamefont
  {O.}~\bibnamefont {Vinyals}}, \bibinfo {author} {\bibfnamefont
  {P.}~\bibnamefont {Warden}}, \bibinfo {author} {\bibfnamefont
  {M.}~\bibnamefont {Wattenberg}}, \bibinfo {author} {\bibfnamefont
  {M.}~\bibnamefont {Wicke}}, \bibinfo {author} {\bibfnamefont
  {Y.}~\bibnamefont {Yu}},\ and\ \bibinfo {author} {\bibfnamefont
  {X.}~\bibnamefont {Zheng}},\ }\href {https://www.tensorflow.org/} {\bibinfo
  {title} {{TensorFlow: Large-Scale Machine Learning on Heterogeneous
  Systems}}} (\bibinfo {year} {2015}),\ \bibinfo {note} {software available
  from tensorflow.org}\BibitemShut {NoStop}%
\bibitem [{\citenamefont {Bradbury}\ \emph {et~al.}(2018)\citenamefont
  {Bradbury}, \citenamefont {Frostig}, \citenamefont {Hawkins}, \citenamefont
  {Johnson}, \citenamefont {Leary}, \citenamefont {Maclaurin}, \citenamefont
  {Necula}, \citenamefont {Paszke}, \citenamefont {Vander{P}las}, \citenamefont
  {Wanderman-{M}ilne},\ and\ \citenamefont {Zhang}}]{jax2018github}%
  \BibitemOpen
  \bibfield  {author} {\bibinfo {author} {\bibfnamefont {J.}~\bibnamefont
  {Bradbury}}, \bibinfo {author} {\bibfnamefont {R.}~\bibnamefont {Frostig}},
  \bibinfo {author} {\bibfnamefont {P.}~\bibnamefont {Hawkins}}, \bibinfo
  {author} {\bibfnamefont {M.~J.}\ \bibnamefont {Johnson}}, \bibinfo {author}
  {\bibfnamefont {C.}~\bibnamefont {Leary}}, \bibinfo {author} {\bibfnamefont
  {D.}~\bibnamefont {Maclaurin}}, \bibinfo {author} {\bibfnamefont
  {G.}~\bibnamefont {Necula}}, \bibinfo {author} {\bibfnamefont
  {A.}~\bibnamefont {Paszke}}, \bibinfo {author} {\bibfnamefont
  {J.}~\bibnamefont {Vander{P}las}}, \bibinfo {author} {\bibfnamefont
  {S.}~\bibnamefont {Wanderman-{M}ilne}},\ and\ \bibinfo {author}
  {\bibfnamefont {Q.}~\bibnamefont {Zhang}},\ }\href
  {http://github.com/google/jax} {\bibinfo {title} {{JAX}: composable
  transformations of {P}ython+{N}um{P}y programs}} (\bibinfo {year}
  {2018})\BibitemShut {NoStop}%
\bibitem [{\citenamefont {Paszke}\ \emph {et~al.}(2019)\citenamefont {Paszke},
  \citenamefont {Gross}, \citenamefont {Massa}, \citenamefont {Lerer},
  \citenamefont {Bradbury}, \citenamefont {Chanan}, \citenamefont {Killeen},
  \citenamefont {Lin}, \citenamefont {Gimelshein}, \citenamefont {Antiga},
  \citenamefont {Desmaison}, \citenamefont {Kopf}, \citenamefont {Yang},
  \citenamefont {DeVito}, \citenamefont {Raison}, \citenamefont {Tejani},
  \citenamefont {Chilamkurthy}, \citenamefont {Steiner}, \citenamefont {Fang},
  \citenamefont {Bai},\ and\ \citenamefont
  {Chintala}}]{NEURIPS2019_9015pytorch}%
  \BibitemOpen
  \bibfield  {author} {\bibinfo {author} {\bibfnamefont {A.}~\bibnamefont
  {Paszke}}, \bibinfo {author} {\bibfnamefont {S.}~\bibnamefont {Gross}},
  \bibinfo {author} {\bibfnamefont {F.}~\bibnamefont {Massa}}, \bibinfo
  {author} {\bibfnamefont {A.}~\bibnamefont {Lerer}}, \bibinfo {author}
  {\bibfnamefont {J.}~\bibnamefont {Bradbury}}, \bibinfo {author}
  {\bibfnamefont {G.}~\bibnamefont {Chanan}}, \bibinfo {author} {\bibfnamefont
  {T.}~\bibnamefont {Killeen}}, \bibinfo {author} {\bibfnamefont
  {Z.}~\bibnamefont {Lin}}, \bibinfo {author} {\bibfnamefont {N.}~\bibnamefont
  {Gimelshein}}, \bibinfo {author} {\bibfnamefont {L.}~\bibnamefont {Antiga}},
  \bibinfo {author} {\bibfnamefont {A.}~\bibnamefont {Desmaison}}, \bibinfo
  {author} {\bibfnamefont {A.}~\bibnamefont {Kopf}}, \bibinfo {author}
  {\bibfnamefont {E.}~\bibnamefont {Yang}}, \bibinfo {author} {\bibfnamefont
  {Z.}~\bibnamefont {DeVito}}, \bibinfo {author} {\bibfnamefont
  {M.}~\bibnamefont {Raison}}, \bibinfo {author} {\bibfnamefont
  {A.}~\bibnamefont {Tejani}}, \bibinfo {author} {\bibfnamefont
  {S.}~\bibnamefont {Chilamkurthy}}, \bibinfo {author} {\bibfnamefont
  {B.}~\bibnamefont {Steiner}}, \bibinfo {author} {\bibfnamefont
  {L.}~\bibnamefont {Fang}}, \bibinfo {author} {\bibfnamefont {J.}~\bibnamefont
  {Bai}},\ and\ \bibinfo {author} {\bibfnamefont {S.}~\bibnamefont
  {Chintala}},\ }\bibfield  {title} {\bibinfo {title} {{PyTorch: An Imperative
  Style, High-Performance Deep Learning Library}},\ }in\ \href
  {http://papers.neurips.cc/paper/9015-pytorch-an-imperative-style-high-performance-deep-learning-library.pdf}
  {\emph {\bibinfo {booktitle} {NeurIPS}}},\ \bibinfo {editor} {edited by\
  \bibinfo {editor} {\bibfnamefont {H.}~\bibnamefont {Wallach}}, \bibinfo
  {editor} {\bibfnamefont {H.}~\bibnamefont {Larochelle}}, \bibinfo {editor}
  {\bibfnamefont {A.}~\bibnamefont {Beygelzimer}}, \bibinfo {editor}
  {\bibfnamefont {F.}~\bibnamefont {d\textquotesingle Alch\'{e}-Buc}}, \bibinfo
  {editor} {\bibfnamefont {E.}~\bibnamefont {Fox}},\ and\ \bibinfo {editor}
  {\bibfnamefont {R.}~\bibnamefont {Garnett}}}\ (\bibinfo  {publisher} {Curran
  Associates, Inc.},\ \bibinfo {year} {2019})\ pp.\ \bibinfo {pages}
  {8024--8035}\BibitemShut {NoStop}%
\bibitem [{\citenamefont {Heinrich}\ \emph {et~al.}()\citenamefont {Heinrich},
  \citenamefont {Feickert},\ and\ \citenamefont {Stark}}]{pyhf}%
  \BibitemOpen
  \bibfield  {author} {\bibinfo {author} {\bibfnamefont {L.}~\bibnamefont
  {Heinrich}}, \bibinfo {author} {\bibfnamefont {M.}~\bibnamefont {Feickert}},\
  and\ \bibinfo {author} {\bibfnamefont {G.}~\bibnamefont {Stark}},\ }\href
  {https://doi.org/10.5281/zenodo.1169739} {\bibinfo {title} {{pyhf:
  v0.6.3}}},\ \bibinfo {note}
  {https://github.com/scikit-hep/pyhf/releases/tag/v0.6.3}\BibitemShut
  {NoStop}%
\bibitem [{\citenamefont {Heinrich}\ \emph {et~al.}(2021)\citenamefont
  {Heinrich}, \citenamefont {Feickert}, \citenamefont {Stark},\ and\
  \citenamefont {Cranmer}}]{pyhf_joss}%
  \BibitemOpen
  \bibfield  {author} {\bibinfo {author} {\bibfnamefont {L.}~\bibnamefont
  {Heinrich}}, \bibinfo {author} {\bibfnamefont {M.}~\bibnamefont {Feickert}},
  \bibinfo {author} {\bibfnamefont {G.}~\bibnamefont {Stark}},\ and\ \bibinfo
  {author} {\bibfnamefont {K.}~\bibnamefont {Cranmer}},\ }\bibfield  {title}
  {\bibinfo {title} {{pyhf: pure-Python implementation of HistFactory
  statistical models}},\ }\href {https://doi.org/10.21105/joss.02823}
  {\bibfield  {journal} {\bibinfo  {journal} {JOSS}\ }\textbf {\bibinfo
  {volume} {6}},\ \bibinfo {pages} {2823} (\bibinfo {year} {2021})}\BibitemShut
  {NoStop}%
\bibitem [{\citenamefont {Heinrich}\ and\ \citenamefont
  {Simpson}(2020)}]{lukas_heinrich_2020_3697981}%
  \BibitemOpen
  \bibfield  {author} {\bibinfo {author} {\bibfnamefont {L.}~\bibnamefont
  {Heinrich}}\ and\ \bibinfo {author} {\bibfnamefont {N.}~\bibnamefont
  {Simpson}},\ }\href {https://doi.org/10.5281/zenodo.3697981} {\bibinfo
  {title} {{pyhf/neos: initial zenodo release
  [\href{https://doi.org/10.5281/zenodo.3697981}{Link}]}}} (\bibinfo {year}
  {2020})\BibitemShut {NoStop}%
\bibitem [{\citenamefont {Carrazza}\ \emph {et~al.}(2021)\citenamefont
  {Carrazza}, \citenamefont {Cruz-Martinez},\ and\ \citenamefont
  {Rossi}}]{pdflow2021}%
  \BibitemOpen
  \bibfield  {author} {\bibinfo {author} {\bibfnamefont {S.}~\bibnamefont
  {Carrazza}}, \bibinfo {author} {\bibfnamefont {J.~M.}\ \bibnamefont
  {Cruz-Martinez}},\ and\ \bibinfo {author} {\bibfnamefont {M.}~\bibnamefont
  {Rossi}},\ }\bibfield  {title} {\bibinfo {title} {Pdfflow: Parton
  distribution functions on gpu},\ }\href
  {https://doi.org/10.1016/j.cpc.2021.107995} {\bibfield  {journal} {\bibinfo
  {journal} {Comp. Phys. Comm.}\ }\textbf {\bibinfo {volume} {264}},\ \bibinfo
  {pages} {107995} (\bibinfo {year} {2021})}\BibitemShut {NoStop}%
\bibitem [{\citenamefont {Ball}\ \emph {et~al.}(2021)\citenamefont {Ball},
  \citenamefont {Carrazza}, \citenamefont {Cruz-Martinez}, \citenamefont
  {Debbio}, \citenamefont {Forte}, \citenamefont {Giani}, \citenamefont
  {Iranipour}, \citenamefont {Kassabov}, \citenamefont {Latorre}, \citenamefont
  {Nocera}, \citenamefont {Pearson}, \citenamefont {Rojo}, \citenamefont
  {Stegeman}, \citenamefont {Schwan}, \citenamefont {Ubiali}, \citenamefont
  {Voisey},\ and\ \citenamefont {Wilson}}]{ball2021opensource}%
  \BibitemOpen
  \bibfield  {author} {\bibinfo {author} {\bibfnamefont {R.~D.}\ \bibnamefont
  {Ball}}, \bibinfo {author} {\bibfnamefont {S.}~\bibnamefont {Carrazza}},
  \bibinfo {author} {\bibfnamefont {J.}~\bibnamefont {Cruz-Martinez}}, \bibinfo
  {author} {\bibfnamefont {L.~D.}\ \bibnamefont {Debbio}}, \bibinfo {author}
  {\bibfnamefont {S.}~\bibnamefont {Forte}}, \bibinfo {author} {\bibfnamefont
  {T.}~\bibnamefont {Giani}}, \bibinfo {author} {\bibfnamefont
  {S.}~\bibnamefont {Iranipour}}, \bibinfo {author} {\bibfnamefont
  {Z.}~\bibnamefont {Kassabov}}, \bibinfo {author} {\bibfnamefont {J.~I.}\
  \bibnamefont {Latorre}}, \bibinfo {author} {\bibfnamefont {E.~R.}\
  \bibnamefont {Nocera}}, \bibinfo {author} {\bibfnamefont {R.~L.}\
  \bibnamefont {Pearson}}, \bibinfo {author} {\bibfnamefont {J.}~\bibnamefont
  {Rojo}}, \bibinfo {author} {\bibfnamefont {R.}~\bibnamefont {Stegeman}},
  \bibinfo {author} {\bibfnamefont {C.}~\bibnamefont {Schwan}}, \bibinfo
  {author} {\bibfnamefont {M.}~\bibnamefont {Ubiali}}, \bibinfo {author}
  {\bibfnamefont {C.}~\bibnamefont {Voisey}},\ and\ \bibinfo {author}
  {\bibfnamefont {M.}~\bibnamefont {Wilson}},\ }\href@noop {} {\bibinfo {title}
  {An open-source machine learning framework for global analyses of parton
  distributions}} (\bibinfo {year} {2021}),\ \Eprint
  {https://arxiv.org/abs/2109.02671} {arXiv:2109.02671 [hep-ph]} \BibitemShut
  {NoStop}%
\bibitem [{\citenamefont {Kagan}\ and\ \citenamefont
  {Heinrich}(2023)}]{kagan2023branches}%
  \BibitemOpen
  \bibfield  {author} {\bibinfo {author} {\bibfnamefont {M.}~\bibnamefont
  {Kagan}}\ and\ \bibinfo {author} {\bibfnamefont {L.}~\bibnamefont
  {Heinrich}},\ }\href@noop {} {\bibinfo {title} {Branches of a tree: Taking
  derivatives of programs with discrete and branching randomness in high energy
  physics}} (\bibinfo {year} {2023}),\ \Eprint
  {https://arxiv.org/abs/2308.16680} {arXiv:2308.16680 [stat.ML]} \BibitemShut
  {NoStop}%
\bibitem [{\citenamefont {Nachman}\ and\ \citenamefont
  {Prestel}(2022)}]{nachman2022morphing}%
  \BibitemOpen
  \bibfield  {author} {\bibinfo {author} {\bibfnamefont {B.}~\bibnamefont
  {Nachman}}\ and\ \bibinfo {author} {\bibfnamefont {S.}~\bibnamefont
  {Prestel}},\ }\href@noop {} {\bibinfo {title} {Morphing parton showers with
  event derivatives}} (\bibinfo {year} {2022}),\ \Eprint
  {https://arxiv.org/abs/2208.02274} {arXiv:2208.02274 [hep-ph]} \BibitemShut
  {NoStop}%
\bibitem [{\citenamefont {Heinrich}\ and\ \citenamefont
  {Kagan}(2023)}]{Heinrich_2023}%
  \BibitemOpen
  \bibfield  {author} {\bibinfo {author} {\bibfnamefont {L.}~\bibnamefont
  {Heinrich}}\ and\ \bibinfo {author} {\bibfnamefont {M.}~\bibnamefont
  {Kagan}},\ }\bibfield  {title} {\bibinfo {title} {Differentiable matrix
  elements with madjax},\ }\href
  {https://doi.org/10.1088/1742-6596/2438/1/012137} {\bibfield  {journal}
  {\bibinfo  {journal} {J. Phy. Conf. Series}\ }\textbf {\bibinfo {volume}
  {2438}},\ \bibinfo {pages} {012137} (\bibinfo {year} {2023})}\BibitemShut
  {NoStop}%
\bibitem [{\citenamefont {Li}\ \emph {et~al.}(2016)\citenamefont {Li},
  \citenamefont {Tarlow}, \citenamefont {Brockschmidt},\ and\ \citenamefont
  {Zemel}}]{DBLP:journals/corr/LiTBZ15}%
  \BibitemOpen
  \bibfield  {author} {\bibinfo {author} {\bibfnamefont {Y.}~\bibnamefont
  {Li}}, \bibinfo {author} {\bibfnamefont {D.}~\bibnamefont {Tarlow}}, \bibinfo
  {author} {\bibfnamefont {M.}~\bibnamefont {Brockschmidt}},\ and\ \bibinfo
  {author} {\bibfnamefont {R.~S.}\ \bibnamefont {Zemel}},\ }\bibfield  {title}
  {\bibinfo {title} {Gated graph sequence neural networks},\ }in\ \href
  {http://arxiv.org/abs/1511.05493} {\emph {\bibinfo {booktitle} {4th
  International Conference on Learning Representations, {ICLR} 2016, San Juan,
  Puerto Rico, May 2-4, 2016, Conference Track Proceedings}}},\ \bibinfo
  {editor} {edited by\ \bibinfo {editor} {\bibfnamefont {Y.}~\bibnamefont
  {Bengio}}\ and\ \bibinfo {editor} {\bibfnamefont {Y.}~\bibnamefont {LeCun}}}\
  (\bibinfo {year} {2016})\BibitemShut {NoStop}%
\bibitem [{\citenamefont {He}\ \emph {et~al.}(2016)\citenamefont {He},
  \citenamefont {Zhang}, \citenamefont {Ren},\ and\ \citenamefont
  {Sun}}]{He_2016_CVPR}%
  \BibitemOpen
  \bibfield  {author} {\bibinfo {author} {\bibfnamefont {K.}~\bibnamefont
  {He}}, \bibinfo {author} {\bibfnamefont {X.}~\bibnamefont {Zhang}}, \bibinfo
  {author} {\bibfnamefont {S.}~\bibnamefont {Ren}},\ and\ \bibinfo {author}
  {\bibfnamefont {J.}~\bibnamefont {Sun}},\ }\bibfield  {title} {\bibinfo
  {title} {Deep residual learning for image recognition},\ }in\ \href@noop {}
  {\emph {\bibinfo {booktitle} {Proceedings of the IEEE Conference on Computer
  Vision and Pattern Recognition (CVPR)}}}\ (\bibinfo {year}
  {2016})\BibitemShut {NoStop}%
\bibitem [{\citenamefont {Ba}\ \emph {et~al.}(2016)\citenamefont {Ba},
  \citenamefont {Kiros},\ and\ \citenamefont {Hinton}}]{ba2016layer}%
  \BibitemOpen
  \bibfield  {author} {\bibinfo {author} {\bibfnamefont {J.~L.}\ \bibnamefont
  {Ba}}, \bibinfo {author} {\bibfnamefont {J.~R.}\ \bibnamefont {Kiros}},\ and\
  \bibinfo {author} {\bibfnamefont {G.~E.}\ \bibnamefont {Hinton}},\
  }\href@noop {} {\bibinfo {title} {Layer normalization}} (\bibinfo {year}
  {2016}),\ \Eprint {https://arxiv.org/abs/1607.06450} {arXiv:1607.06450
  [stat.ML]} \BibitemShut {NoStop}%
\bibitem [{\citenamefont {Ginsburg}\ \emph {et~al.}(2020)\citenamefont
  {Ginsburg}, \citenamefont {Castonguay}, \citenamefont {Hrinchuk},
  \citenamefont {Kuchaiev}, \citenamefont {Lavrukhin}, \citenamefont {Leary},
  \citenamefont {Li}, \citenamefont {Nguyen}, \citenamefont {Zhang},\ and\
  \citenamefont {Cohen}}]{novograd}%
  \BibitemOpen
  \bibfield  {author} {\bibinfo {author} {\bibfnamefont {B.}~\bibnamefont
  {Ginsburg}}, \bibinfo {author} {\bibfnamefont {P.}~\bibnamefont
  {Castonguay}}, \bibinfo {author} {\bibfnamefont {O.}~\bibnamefont
  {Hrinchuk}}, \bibinfo {author} {\bibfnamefont {O.}~\bibnamefont {Kuchaiev}},
  \bibinfo {author} {\bibfnamefont {V.}~\bibnamefont {Lavrukhin}}, \bibinfo
  {author} {\bibfnamefont {R.}~\bibnamefont {Leary}}, \bibinfo {author}
  {\bibfnamefont {J.}~\bibnamefont {Li}}, \bibinfo {author} {\bibfnamefont
  {H.}~\bibnamefont {Nguyen}}, \bibinfo {author} {\bibfnamefont
  {Y.}~\bibnamefont {Zhang}},\ and\ \bibinfo {author} {\bibfnamefont {J.~M.}\
  \bibnamefont {Cohen}},\ }\href@noop {} {\bibinfo {title} {Stochastic gradient
  methods with layer-wise adaptive moments for training of deep networks}}
  (\bibinfo {year} {2020}),\ \Eprint {https://arxiv.org/abs/1905.11286}
  {arXiv:1905.11286 [cs]} \BibitemShut {NoStop}%
\bibitem [{\citenamefont {Shlomi}(2020)}]{zenodo_dataset}%
  \BibitemOpen
  \bibfield  {author} {\bibinfo {author} {\bibfnamefont {J.}~\bibnamefont
  {Shlomi}},\ }\bibfield  {title} {\bibinfo {title} {Secondary vertex finding
  in jets dataset},\ }\href {https://doi.org/10.5281/zenodo.4044628}
  {10.5281/zenodo.4044628} (\bibinfo {year} {2020})\BibitemShut {NoStop}%
\bibitem [{\citenamefont {Sj{\"o}strand}\ \emph {et~al.}(2008)\citenamefont
  {Sj{\"o}strand}, \citenamefont {Mrenna},\ and\ \citenamefont
  {Skands}}]{Sjostrand:2007gs}%
  \BibitemOpen
  \bibfield  {author} {\bibinfo {author} {\bibfnamefont {T.}~\bibnamefont
  {Sj{\"o}strand}}, \bibinfo {author} {\bibfnamefont {S.}~\bibnamefont
  {Mrenna}},\ and\ \bibinfo {author} {\bibfnamefont {P.}~\bibnamefont
  {Skands}},\ }\bibfield  {title} {\bibinfo {title} {{A brief introduction to
  PYTHIA 8.1}},\ }\href {https://doi.org/10.1016/j.cpc.2008.01.036} {\bibfield
  {journal} {\bibinfo  {journal} {Comput. Phys. Commun.}\ }\textbf {\bibinfo
  {volume} {178}},\ \bibinfo {pages} {852} (\bibinfo {year} {2008})},\ \Eprint
  {https://arxiv.org/abs/0710.3820} {arXiv:0710.3820 [hep-ph]} \BibitemShut
  {NoStop}%
%%CITATION = 0710.3820;%%
\bibitem [{\citenamefont {de~Favereau}\ \emph {et~al.}(2014)\citenamefont
  {de~Favereau}, \citenamefont {Delaere}, \citenamefont {Demin}, \citenamefont
  {Giammanco}, \citenamefont {Lema\^\i{}tre}, \citenamefont {Mertens},\ and\
  \citenamefont {Selvaggi}}]{deFavereau:2013fsa}%
  \BibitemOpen
  \bibfield  {author} {\bibinfo {author} {\bibfnamefont {J.}~\bibnamefont
  {de~Favereau}}, \bibinfo {author} {\bibfnamefont {C.}~\bibnamefont
  {Delaere}}, \bibinfo {author} {\bibfnamefont {P.}~\bibnamefont {Demin}},
  \bibinfo {author} {\bibfnamefont {A.}~\bibnamefont {Giammanco}}, \bibinfo
  {author} {\bibfnamefont {V.}~\bibnamefont {Lema\^\i{}tre}}, \bibinfo {author}
  {\bibfnamefont {A.}~\bibnamefont {Mertens}},\ and\ \bibinfo {author}
  {\bibfnamefont {M.}~\bibnamefont {Selvaggi}} (\bibinfo {collaboration}
  {DELPHES 3}),\ }\bibfield  {title} {\bibinfo {title} {{DELPHES 3, A modular
  framework for fast simulation of a generic collider experiment}},\ }\href
  {https://doi.org/10.1007/JHEP02(2014)057} {\bibfield  {journal} {\bibinfo
  {journal} {JHEP}\ }\textbf {\bibinfo {volume} {02}},\ \bibinfo {pages}
  {057}},\ \Eprint {https://arxiv.org/abs/1307.6346} {arXiv:1307.6346 [hep-ex]}
  \BibitemShut {NoStop}%
\bibitem [{\citenamefont {Cacciari}\ \emph {et~al.}(2008)\citenamefont
  {Cacciari}, \citenamefont {Salam},\ and\ \citenamefont {Soyez}}]{antikt}%
  \BibitemOpen
  \bibfield  {author} {\bibinfo {author} {\bibfnamefont {M.}~\bibnamefont
  {Cacciari}}, \bibinfo {author} {\bibfnamefont {G.~P.}\ \bibnamefont
  {Salam}},\ and\ \bibinfo {author} {\bibfnamefont {G.}~\bibnamefont {Soyez}},\
  }\bibfield  {title} {\bibinfo {title} {The anti-kt jet clustering
  algorithm},\ }\href {https://doi.org/10.1088/1126-6708/2008/04/063}
  {\bibfield  {journal} {\bibinfo  {journal} {Journal of High Energy Physics}\
  }\textbf {\bibinfo {volume} {2008}},\ \bibinfo {pages} {063} (\bibinfo {year}
  {2008})}\BibitemShut {NoStop}%
\end{thebibliography}%

\clearpage
\appendix

\section{Generalized vertex fit formulated as a nonlinear regression}\label{app:vertexfit}

We review here the inclusive vertex fitting strategy of the Billoir Algorithm~\cite{BILLOIR1985115}. Note that while our treatment includes weights $\{w\}$, for the derivation of the solution those weights can effectively be absorbed into the covariance matrix, thus resulting in the same derivation.

Suppose that we have a set of $N$ tracks that we want to fit to a single vertex. These tracks are specified by the perigee parameterization with respect to the primary vertex such that the $i$th track has track parameters $\mathbf{q}_i = (d_0, z_0, \phi, \theta, \rho)$ with associated covariance matrix $\mathbf{V}_i$ where $i = 1, \ldots ,N$. These parameters are defined as follows:

\begin{itemize}
\item $d_0$: signed transverse impact parameter
\item $z_0$: longitudinal impact parameter
\item $\phi$: polar angle of trajectory
\item $\theta$: azimuthal angle of trajectory
\item $\rho$: signed curvature
\end{itemize}

These track parameters $\mathbf{q}_i$ are defined as the nonlinear function of the vertex position $\mathbf{v} = (x_v, y_v, z_v)$ and the momentum vectors of the tracks at that position \mbox{$\mathbf{p}_i = (\theta, \phi_v, \rho)$}.
\begin{equation}
    \mathbf{q}_i = \mathbf{h}_i(\mathbf{v},\mathbf{p}_i), \ \ i = 1, \ldots ,N 
\end{equation}
If we introduce the quantities $Q = x_v \cos(\phi_v) + y_v \sin(\phi_v)$ and $R = y_v \cos(\phi_v) - x_v \sin(\phi_v)$ we can write this calculation at first order in $\rho$ (note that $\theta$ and $\rho$ do not change going from the vertex to the perigee):
\begin{equation}
    \left\{ \begin{aligned} 
        d_0 &= -R - Q^2 \rho /2\\
        z_0 &= z_v - Q (1 - R \rho) \cot(\theta) \\
        \phi &= \phi_v - Q \rho
    \end{aligned}\right.
\end{equation}
Next we calculate the first-order Taylor expansion of $\mathbf{h}_i$, expanded at an estimate of the vertex position and vertex track momenta $\mathbf{e}_0 = (\mathbf{v}_0, \mathbf{p}_{i,0})$
\begin{equation}
    \mathbf{q}_i \approx \mathbf{A}_i \mathbf{v} + \mathbf{B}_i \mathbf{p}_i + \mathbf{c}_i, \ \ i = 1, \ldots ,N
\end{equation}
with
\begin{equation}
    \begin{aligned} 
        \mathbf{A}_i &= \left. \dfrac{\partial \mathbf{h}_i}{\partial \mathbf{v}} \right\vert_{\mathbf{e}_0} = \begin{pmatrix} s & -c & 0 \\ -tc & -ts & 1 \\ -\rho c & -\rho s & 0 \\ 0 & 0 & 0 \\ 0 & 0 & 0 \end{pmatrix} \\
        \mathbf{B}_i &= \left. \dfrac{\partial \mathbf{h}_i}{\partial \mathbf{p}_i} \right\vert_{\mathbf{e}_0} = \begin{pmatrix} 0 & Q & -Q^2/2 \\ Q (1 + t^2) & -Rt & QRt \\ 0 & 1 & -Q \\ 1 & 0 & 0 \\ 0 & 0 & 1 \end{pmatrix} \\
        \mathbf{c}_i &= \mathbf{h}_i (\mathbf{v}_0, \mathbf{p}_{i,0}) - \mathbf{A}_i \mathbf{v}_0 - \mathbf{B}_i \mathbf{p}_{i,0}
    \end{aligned}
\end{equation}
where $c \equiv \cos(\phi_v)$, $s \equiv \sin(\phi_v)$, and $t \equiv \cot(\theta)$.

From here we compute the following quantities 
\begin{eqnarray}
    \mathbf{G}_i &=& \mathbf{V}_i^{-1} \\
    \mathbf{D}_i &=& \mathbf{A}_i^T \mathbf{G}_i \mathbf{B}_i \\
    \mathbf{D}_0 &=& \sum_{i=1}^N \mathbf{A}_i^T \mathbf{G}_i \mathbf{A}_i \\
    \mathbf{W}_i^{-1} &=& \mathbf{B}_i^T \mathbf{G}_i \mathbf{B}_i
\end{eqnarray}
\begin{equation}
    \mathbf{C} = \left( \mathbf{D}_0 - \sum_{i=1}^N \mathbf{D}_i \mathbf{W}_i \mathbf{D}_i^T \right)^{-1}
\end{equation}
which are then used in the calculation of the estimated vertex parameters:
\begin{equation}
    \hat{\mathbf{v}} = \mathbf{C} \sum_{i=1}^N \mathbf{A}_i^T \mathbf{G}_i (\mathbf{I} - \mathbf{B}_i \mathbf{W}_i \mathbf{B}_i^T \mathbf{G}_i)(\mathbf{q}_i - \mathbf{c}_i)
\end{equation}
\begin{equation}
    \hat{\mathbf{p}}_i = \mathbf{W}_i \mathbf{B}_i^T \mathbf{G}_i (\mathbf{q}_i - \mathbf{c}_i - \mathbf{A}_i \hat{\mathbf{v}}), \ \ i = 1, \ldots ,N
\end{equation}

This fit is then iterated until convergence, expanding the functions $\mathbf{h}_i$ around the new expansion point $\mathbf{e} = (\hat{\mathbf{v}}, \hat{\mathbf{p}}_i)$ each time. Afterwards we rewrite the track parameters:
\begin{equation}
    \hat{\mathbf{q}}_i = \mathbf{h}_i (\hat{\mathbf{v}}, \hat{\mathbf{p}}_i), \ \ i = 1, \ldots ,N
\end{equation}
The $\chi^2$ statistic of the fit is then:
\begin{equation}
    \chi^2 = \sum_{i=1}^N (\mathbf{q}_i - \hat{\mathbf{q}}_i)^T \mathbf{G}_i (\mathbf{q}_i - \hat{\mathbf{q}}_i)
\end{equation}

\setcounter{figure}{10}
\begin{figure*}[ht!]
\centering
    \begin{subfigure}[b]{0.3\textwidth}
    \centering
    \includegraphics[width=1\textwidth]{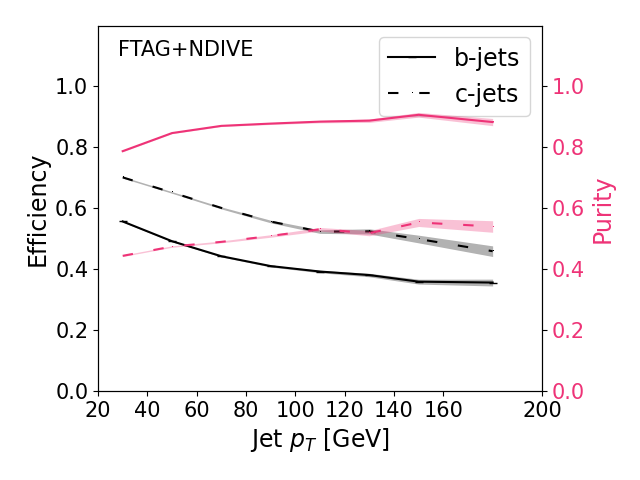}
    \end{subfigure}
    \begin{subfigure}[b]{0.3\textwidth}
    \centering
    \includegraphics[width=1\textwidth]{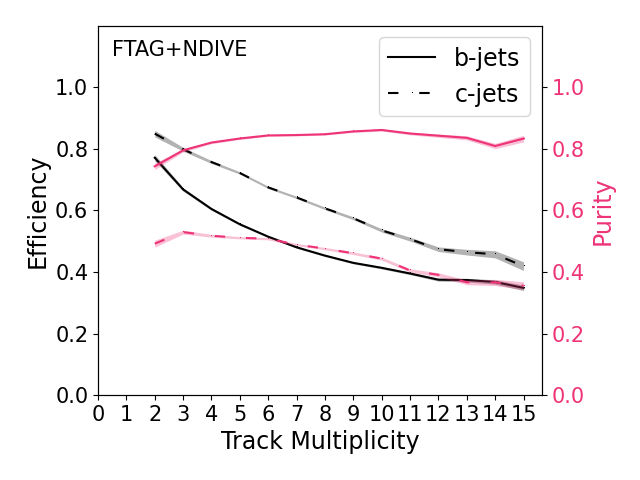}
    \end{subfigure}    
    \begin{subfigure}[b]{0.3\textwidth}
    \centering
    \includegraphics[width=1\textwidth]{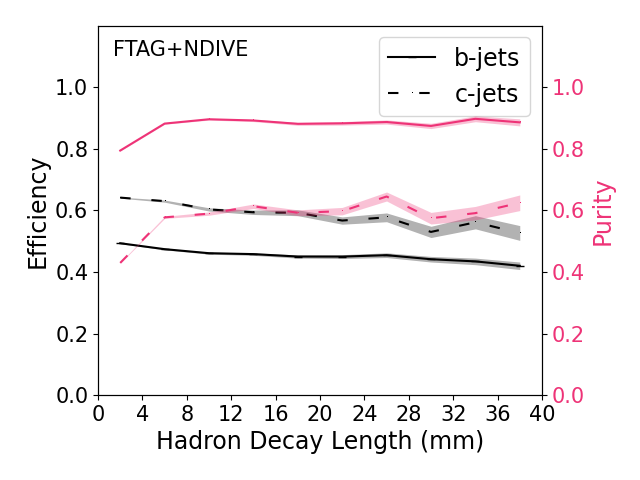}
    \end{subfigure}        
\caption{Track selection efficiencies (left) and purities (right) achieved by \fndive\ with a selection threshold of 0.5, as a function of the jet \pt, track multiplicity, and hadron decay length, for $b$- and $c$-jets. The error bars correspond to the square root of the variance of a binomial distribution.}
\label{fig:ftagndive_eff_purity}
\end{figure*}

\setcounter{figure}{8}
\begin{figure}[ht!]
\centering
    \begin{subfigure}[b]{0.45\textwidth}
    \centering
    \includegraphics[width=0.75\textwidth]{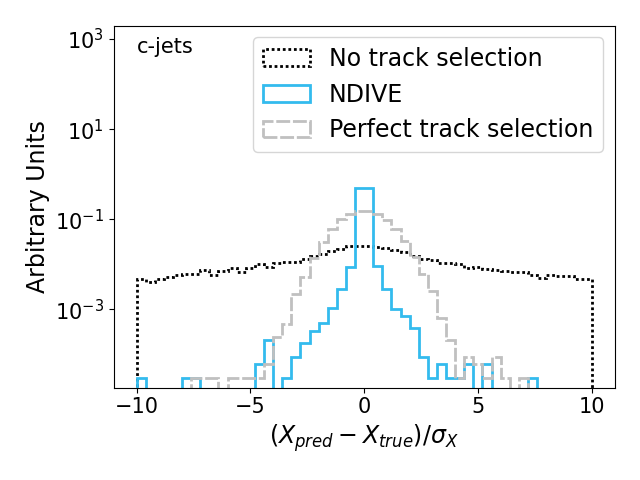}
    \end{subfigure}\\
    \begin{subfigure}[b]{0.45\textwidth}
    \centering
    \includegraphics[width=0.75\textwidth]{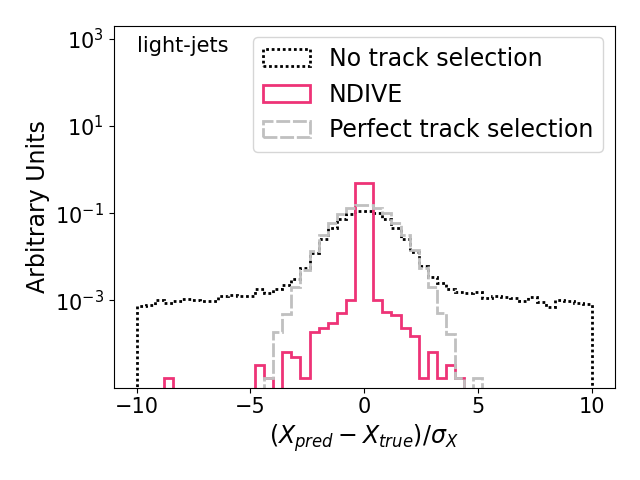}
    \end{subfigure}\\    
\caption{Difference between the fit and true vertex $x$-coordinate divided by the square root of the fit vertex variance, for $c$- and light-jets, comparing \ndive~with vertex fitting with no track selection and with perfect track selection.}
\label{fig:ndive_score}
\end{figure}

\begin{figure}[ht!]
    \centering
    \begin{subfigure}[b]{0.4\textwidth}
    \centering
    \includegraphics[width=\textwidth]{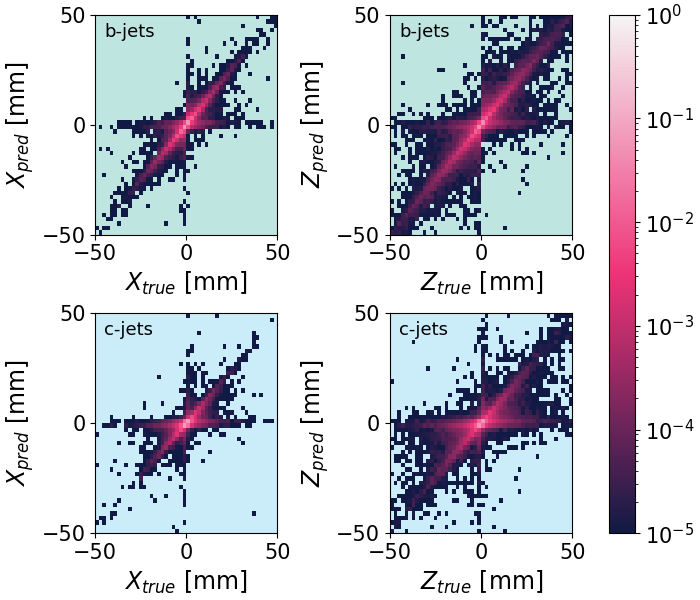}
    \end{subfigure}\\
\caption{Fitted and true vertex coordinates $x$ and $z$ for $b$-jets (top) and $c$-jets (bottom).}
\label{fig:fit2D}
\end{figure}

\section{Track Extrapolation}
\label{app:extrap}

We parameterize a generic position $V$ along a track trajectory by considering the track's perigee representation with respect to $P_0=(x_P,y_P,z_P)$ and a scanning parameter $\phi_V$, with which we can express the spatial coordinates of $V$ as~\cite{Giacinto2010}:

\begin{eqnarray}
\nonumber x_V &=& x_P + d_0 \cos \left( \phi + \frac{\pi}{2} \right) \\
              & & + \rho \left[ \cos \left( \phi_V + \frac{\pi}{2} \right) -\cos \left( \phi + \frac{\pi}{2}\right) \right] \\
\nonumber y_V &=& y_P + d_0 \sin \left( \phi + \frac{\pi}{2} \right) \\ 
              & & + \rho \left[ \sin \left( \phi_V + \frac{\pi}{2} \right) -\sin \left( \phi + \frac{\pi}{2}\right) \right] \\
z_V &=& z_P + z_0 - \frac{\rho}{\tan(\theta)} \left[\phi_V - \phi\right]
\end{eqnarray}

Additional track representations can be defined by considering alternative reference points. In order to find the point of closest approach with respect to a new reference point $P^\prime=(x_{P^\prime},y_{P^\prime},z_{P^\prime})$, the three-dimensional euclidean distance $D(z_V)$ between any point along the trajectory $(x_V,y_V,z_V)$ and $P^\prime$ can be calculated and its minimum found (e.g. via a parametric scan of the trajectory around $z_V=z_{P^\prime}$). 

\section{Additional \ndive\ Performance Studies}
\label{app:ndiveperf}

Fig.~\ref{fig:ndive_score} shows the difference between the fit vertex and the true vertex $x$-coordinate, divided by their corresponding standard deviations, for $c$-jets and light-jets. 

Fig.~\ref{fig:fit2D} shows the true vertex versus fit vertex $x$- and $z$- coordinates for $b$-jets and $c$-jets. In general the fit results in a strong correlation between true and fit coordinate values, as indicated by the diagonal distributions. The correlation tends to be slightly lower in $c$-jets, as fewer tracks are available in $c$-jet decays for fitting the vertex. The vertical and horizontal contributions are a symptom of \ndive\ incorrectly attempting to fit either the primary vertex or a spurious vertex.

\section{Additional \fndive\ Performance Studies}
\label{app:fndiveperf}

The efficiency and purity of the track selection in $b$-jets and $c$-jets for the \ndive\ vertex fitter when trained as a part of \fndive\ is shown in Fig.~\ref{fig:ftagndive_eff_purity}. Both are found to be relatively stable as a function of the jet \pt, track multiplicity, and hadron decay length. Both the efficiency and purity are higher in vertex fits from \fndive\ than standalone \ndive.

\end{document}